\documentclass[aps,showpacs,floatfix,nofootinbib,12pt]{revtex4}
\usepackage{graphicx}
\usepackage{epsf}
\usepackage{wrapfig}
\usepackage{epsfig}
\usepackage{sublabel}
\usepackage{epsfig}
\usepackage{verbatim}
\usepackage{amsmath}
\usepackage{color}
\usepackage{soul}
\usepackage{ulem}
\usepackage{multirow}
\def\bk{{\mbox{\boldmath$k$}}}

\def\b0{{\mbox{\boldmath$0$}}}

\def\bk{{\mbox{\boldmath$k$}}}

\def\Vec#1{\mbox{\boldmath $#1$}}

\def\beq{\begin{equation}}
\def\eeq{\end{equation}}

\def\beqy{\begin{eqnarray}}
\def\eeqy{\end{eqnarray}}

\def \b #1{ {\bf #1}}
\newcommand{\be}{\begin{eqnarray}}
\newcommand{\ee}{\end{eqnarray}}

\def \b #1{ {\bf #1}}

\def \b #1{ {\bf #1}}
     \font\tenbifull=cmmib10 scaled 1200 
     \font\tenbimed=cmmib9
     \font\tenbismall=cmmib7
       \def\bmit{\fam9 }
\mathchardef\bbkappa="7114 \mathchardef\bbrho="711A
\mathchardef\bbsigma="711B \mathchardef\bbtau="711C
\mathchardef\bbvarrho="7125 \mathchardef\bbvarsigma="7126
\mathchardef\bbxi="7118

\def\boldrho{{\bmit\bbrho}}

\begin{document}
\vskip 2mm
\date{\today}\vskip 2mm
\title{A microscopic nucleon spectral function for finite nuclei
featuring
    two- and three- nucleon short-range correlations:
    The model {\it vs} ab-initio calculations for the three-nucleon systems}
\author{Claudio Ciofi degli Atti$^{1}$}
\email{ciofi@pg.infn.it}
\author{Chiara Benedetta Mezzetti$^{2}$}
\email{<chiara.mezzetti@unipg.it>}
\author{\mbox{Hiko Morita$^3$}}
\email{hiko@webmail.sgu.ac.jp} \affiliation{ $^1$Istituto
Nazionale di Fisica Nucleare,
 Sezione di Perugia,\\
 c/o Department of Physics and Geology, University of Perugia,
  Via A. Pascoli, I-06123, Perugia, Italy\\
$^2$NiPS Laboratory, Department of Physics and Geology, University
of Perugia,
  Via A. Pascoli, I-06123, Perugia, Italy\\
$^3$\mbox{Sapporo Gakuin University, Bunkyo-dai 11, Ebetsu
069-8555, Hokkaido, Japan}} \vskip 2mm
\begin{abstract}

\noindent{\bf Background}:
    Two-nucleon (2N) short-range correlations (SRC) in nuclei have been recently thoroughly  investigated, both theoretically
    and experimentally  and the study  of three-nucleon (3N) SRC, which
    could provide important information on short-range hadronic structure, is underway. Novel  theoretical ideas  concerning 2N and 3N
    SRC are put forward in the present paper.

\noindent {\bf Purpose}: The general features of a microscopic one-nucleon spectral function which includes the effects of both
      2N and 3N SRC
    and its comparison with {\it ab-initio} spectral functions of the three-nucleon systems are illustrated.

    \noindent{\bf Methods}: A microscopic and parameter-free one-nucleon spectral function  expressed
    in terms of a  convolution integral involving  {\it ab-initio} relative and
    center-of-mass
    (c.m.)
    momentum distributions of a 2N pair and
    aimed at describing two- and three nucleon short-range correlations, is obtained by using :
    (i) the   two-nucleon momentum distributions obtained within {\it ab initio} approaches based upon
    nucleon-nucleon interactions of the Argonne family; (ii)   the exact relation between
    one- and two- nucleon momentum distributions; (iii)   the
    fundamental property of factorization of the nuclear wave function at short inter-nucleon  ranges.

\noindent {\bf Results}: The comparison between the  {\it ab-initio}  spectral function  of $^3$He and the one based upon the convolution
integral, shows  that when the latter contains only  two-nucleon short-range correlations
   the removal energy location of the
peaks  and the region around them   exhibited by the {\it ab-initio} spectral function  are correctly predicted, unlike the case of
 the high and low  removal energy tails;  the inclusion of the effects of
    three-nucleon correlations
     brings  the convolution model spectral function  in much better agreement with the {\it ab initio} one;
     it is also found that whereas the
      three-nucleon short-range correlations  dominate
     the high energy removal energy tail
     of the spectral function, their effects on the one-nucleon momentum distribution are  almost
     one order of magnitude less than the effect of two nucleon short
    -range correlations.

     \noindent {\bf Conclusions}: The convolution model of the spectral function of the three-nucleon systems
     featuring both two-and three-nucleon short-range correlations and correctly depending upon the {\it ab initio} two-nucleon relative
     and center-of-mass momentum distributions  provides in the correlation region a satisfactory approximation of the
     spectral function in a wide range of momentum and removal energy. The extension of the model to
     complex nuclei is expected  to  provide a
     realistic microscopic parameter-free model of the spectral function, whose  properties are therefore
      governed by the features of realistic two-nucleon interactions and  the  momentum distributions  in a given nucleus.
\end{abstract}
 \maketitle
\newpage
%
%
\section{Introduction}
The long-standing problem of the role played by  short-range correlations (SRC) in atomic nuclei   has been the object of intense activity
in recent years  both from the theoretical and the experimental points of view (see the review papers given in Ref.\cite{Theor2N}). The
experimental investigation of two-nucleon (2N) SRC has reached high level  of sophistication \cite{Exp_2N} and, at the same time, a series
of theoretical papers, based upon different approaches, have clarified, both qualitatively and quantitatively, the role played by
 SRC in nuclei \cite{Wiringa:2013ala,New_Papers_SRC,CiofidegliAtti:1995qe,Weiss:2015mba,Newpaper,CiofidegliAtti:2010xv,Baldo:1900zz}. In particular, it has  been demonstrated
 (see e.g.
 \cite{CiofidegliAtti:1995qe,Weiss:2015mba,Newpaper,CiofidegliAtti:2010xv,Baldo:1900zz})
that 2N SRC arise from a
 universal and fundamental  property of the nuclear wave function at short inter-nucleon distances,
 namely its factorization  into  a  wave function describing the relative motion of a nucleon pair
 and a function
describing the motion of the center-of- mass (c.m.)  of the pair with respect to the \lq \lq spectator\rq \rq $(A-2)$-nucleon system.
Concerning the role of possible 3N SRC, although important contributions  have already appeared (see e.g. \cite{Frankstr_1,Egiyan,Fomin}),
much remains to be done in order to fully understand their structure and their effects on other relevant nuclear quantities like, e.g., the
one-nucleon momentum distributions and spectral function (SF). It is the aim of this paper to  illustrate a realistic many-body approach to
 the effects of 2N and 3N SRC on the one-nucleon hole spectral
function  and momentum distributions, two quantities  which play a primary role in the study of short-range effects in nuclei. Preliminary
results along the line presented in this paper have been previously given in Ref. \cite{CBM3N}.
%
\section{The definition of the nucleon spectral function and its  description  in the SRC region by
the  convolution model}
\subsection{The nucleon hole spectral function}
As  is well known,  the nucleon (N) hole spectral function $P_A^N({\bf k}_1,E)$ represents the
 joint probability that when the  nucleon \lq \lq N \rq \rq
 (usually called the {\it active nucleon})  with momentum ${\bf k}_1$ is removed
 instantaneously
 from the ground state of the nucleus A, the nucleus $(A-1)$ (usually called
  the {\it spectator nucleus}) is left in
 the excited state
 $E_{A-1}^*=E-E_{min}$,  where $E$ is the so called {\it removal energy}  and
 $E_{min}= M_{A-1}+m_N-M_A=|E_A|-|E_{A-1}|$,  with $E_A$ and $E_{A-1}$   being
  the (negative)
ground-state energy of nuclei $A$ and $A-1$, respectively. The hole spectral function,  which takes into account the fact that nucleons in
nuclei have not only a momentum distribution, but also a distribution in energy,  is trivially related to a well defined many-body
quantity, namely the two-points Green's function (see e.g. \cite{2:Dickoff}). In this paper we use  the following well-known representation
of the  SF $P_A^N( {\bf k}_1 , E )$, namely
\begin{eqnarray}
   P_A^N( {\bf k}_1 , E ) & = &
      \frac{1}{2 J + 1}
      \sum_{M , \sigma_1}
      \langle
         \Psi_A^{JM}| a_{{\bf k}_1\sigma_1}^{\dag}
         \delta \left ( E - ( {\hat H_A} - E_{A}) \right )  a_{{\bf k}_1\sigma_1}|\Psi_A^{JM}
      \rangle \\
   & = &
      \frac{1}{2 J + 1}
      \sum_{M , \sigma_1}
      \sum\hspace{-0.5cm}\int_f
      \left|
         \langle
            \Psi_{A - 1}^f | a_{{\bf k}_1\sigma_1} | \Psi_{JM}^{A}
         \rangle
      \right|^{2}
      \delta \left ( E - ( E_{A - 1}^{f} - E_{A} ) \right )  \\
   & = &
      \frac{1}{2 J + 1} ( 2 \pi )^{- 3}
      \sum_{M, \sigma_1}\,
      \sum \hspace{-0.5cm}\int_f
      \left|
         \int \mathrm{d} {\bf r}_1
            e^{\mathrm{i} \Vec{k}_1 \cdot \Vec{r}_1}
           \,G_{f}^{M \sigma_1} ( {\bf r}_1 )
      \right|^{2}
      \delta \left ( E - ( E_{A - 1}^{f} - E_{A} ) \right ) ,
   \label{eq6-1}
\end{eqnarray}
where  $a_{{\bf k}_1\sigma_1}^{\dag}$ ($a_{{\bf k}_1\sigma_1}$) is the creation (annihilation) operator of a nucleon with momentum ${\bf
k}_1$ and spin $\sigma$, ${\hat H}_A$ is the intrinsic Hamiltonian for A interacting nucleons, and
 the quantity
%
\begin{equation}
   G_{f}^{M \sigma_1} ( \Vec{r}_1 ) =
      \langle
         \chi _{\sigma_1}^{1/2} ,
         \Psi_{A - 1}^{f} (\{ \Vec{x}\}_{A-1} ) |
         \Psi_{A}^{JM} ( {\bf r}_1, \{ \Vec{x}\}_{A-1})\rangle,
   \label{eq6-2}
\end{equation}
which  has been obtained using the completeness relation for the eigenstates
 of the nucleus $(A-1)$
 ($\sum_f |\Psi_{A-1}^f\rangle \langle\Psi_{A-1}^f| =
1$), is the overlap integral between the ground state wave
function of nucleus $A$,
 $\Psi_A^{JM}$, and the wave functions of
the discrete and all possible continuum eigenfunctions
  $\Psi_{A - 1}^{f}$ (with eigenvalue $E_{A - 1}^{f}$ = $E_{A - 1}$ + $E_{A -
1}^{f *}$) of  the nucleus  $(A - 1)$;  eventually, $\{{\bf x}\}$ denotes the set of spin-isospin and radial coordinates. In what follows
the angle integrated SF is normalized according to (${\bf k}_1\equiv{\bf k}, |{\bf k}| \equiv k$)
\begin{equation}
 4\,\pi\, \int P_A^N( k, E )\,k^2\,d\,{ k} \mathrm{d}E = 1.
   \label{eq6-3}
\end{equation}
and the momentum distribution (normalized to one) is linked to the
SF by the {\it momentum sum rule}
\begin{equation}
 \int P_A^N( k, E )\,\mathrm{d}E = n_A^N(k).
   \label{eq6-3a}
\end{equation}
Thanks to its  very definition, the SF  can be represented in the following useful form \cite{CiofidegliAtti:1995qe}
\begin{equation}
   P_A^N( {k}, E ) =
      P_{\mathrm{0}}^N (k, E ) + P_{1}^N (k, E )\ .
\label{eq6-4}
\end{equation}
\noindent where  $P_{\mathrm{0}}^N$ describes the shell-model part (with occupation probability of shell-model states less than one because
of SRC  populating the states above the Fermi level)
\begin{equation}
P^N_{0} ( k, E )= ( 2 \pi )^{- 3} ( 2 J + 1 )^{- 1} \sum_{M,\sigma,f\leq F}
 \left|\int e^{i{\bf k}_1 \cdot {\bf r}_1}G_{f}^{M \sigma}
 ({\bf r}_1 )\ \mathrm{d}{\bf r}_1
 \right|^{2}\delta (E-E_{min}),
 \label{eq6-5}
 \end{equation}
 and  $P_{\mathrm{1}}^N$ describes the contribution from
the discrete and continuum states above the Fermi level originating from
 ground-state  SRC
\begin{equation}
            P^N_{1} (k, E )=
         ( 2 \pi )^{- 3} ( 2 J + 1 )^{- 1}
         \sum_{M , \sigma} \sum_{f>F}\hspace{-0.5cm}\int
            \left|
               \int e^{\mathrm{i} {\bf k}_1 \cdot {\bf r}_1}
                  G_{f}^{M \sigma} ( {\bf r}_1 )\ \mathrm{d}\vec{r}_1
            \right|^{2}\delta (E-E_{A-1}^f).
            \label{eq6-6}
            \end{equation}

\subsection{The {\it ab initio} spectral function of $^3$He:
the Plane Wave Impulse Approximation (PWIA) {\it vs} the Plane
Wave Approximation (PWA)}
\begin{figure} [!ht]
\includegraphics[width=1.0\textwidth,keepaspectratio]{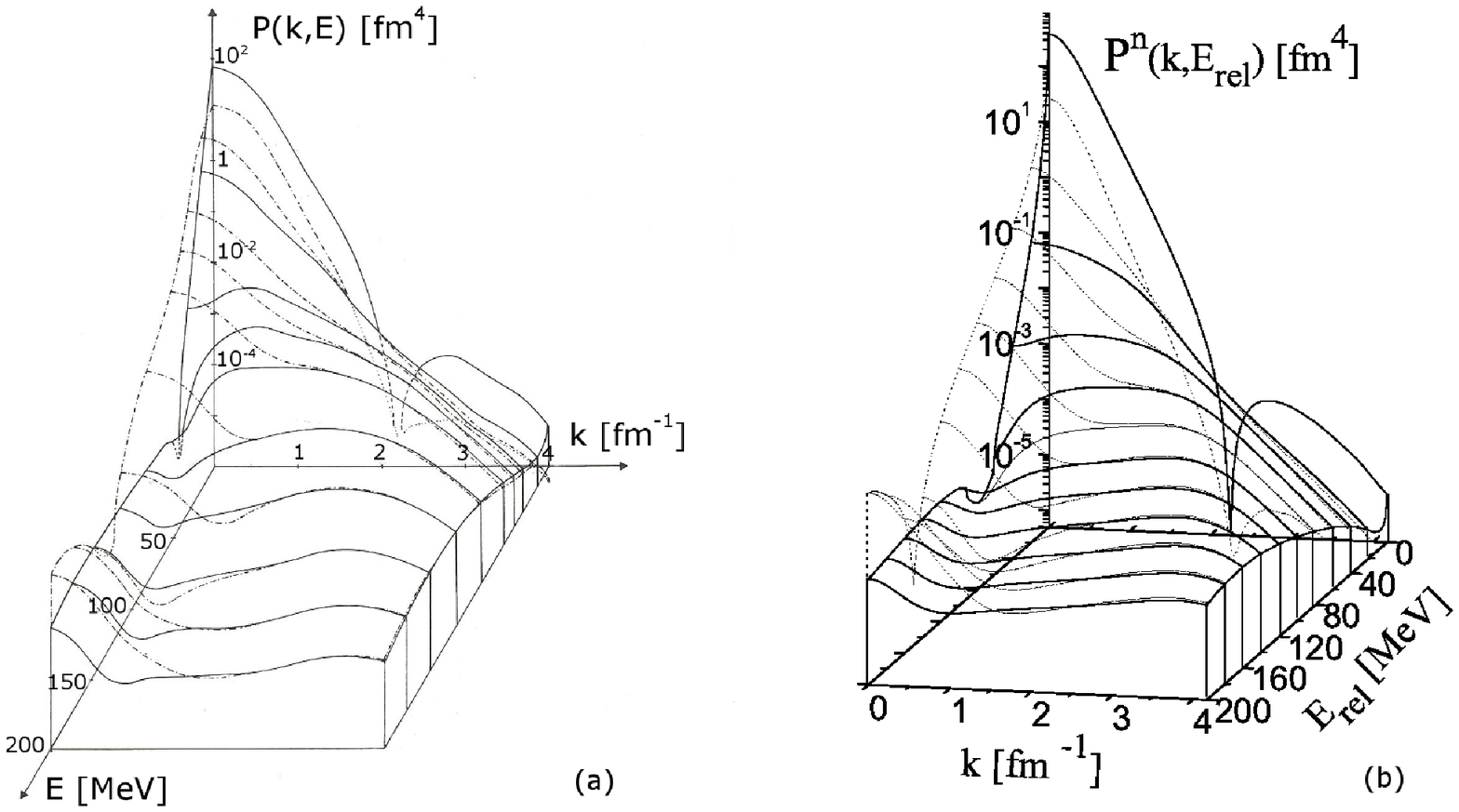}
\caption{(Color online) The {\it ab initio} neutron  spectral
function of $^3$He (Eq. (\ref{piex})) calculated in Ref.
\cite{CPS} using 3N wave functions corresponding to the RSC interaction (\cite{Reid}) (a) and in Ref.
\cite{Ciofi_Kapta} using 3N wave functions \cite{Kievsky:1992um} corresponding to
the AV18 \cite{Wiringa:1994wb}
interaction (b). In both cases the full lines represent the PWIA
(the proton-proton wave function $\Psi_{pp}^{\bf t}$ in the
final state is the exact solution of the
 the Hamiltonian which has been used to obtain the ground-state wave
 function), whereas
 the dot-dashed line in (a)  and the dotted line in (b) correspond
 to the PWA (the
final $pp$ state is approximated by a plane wave). The regions
where the PWA practically coincides with the PWIA are clearly
visible and correspond to high values of the momentum.
In  Figure (b)  $E_{rel}=E+ |E_{3}|$,
$|E_3|$ being the binding energy of
 $^3$He and $E_{rel}$ is the relative energy of the proton-proton pair in the continuum. Therefore the energy scales
 in the two Figures differ  only by the small value  $|E_3|= 7.718 \, MeV$.}
\label{Fig1}
\end{figure}

\begin{figure}[ht]
\begin{center}
\centerline{\includegraphics[width=0.48\textwidth,keepaspectratio]{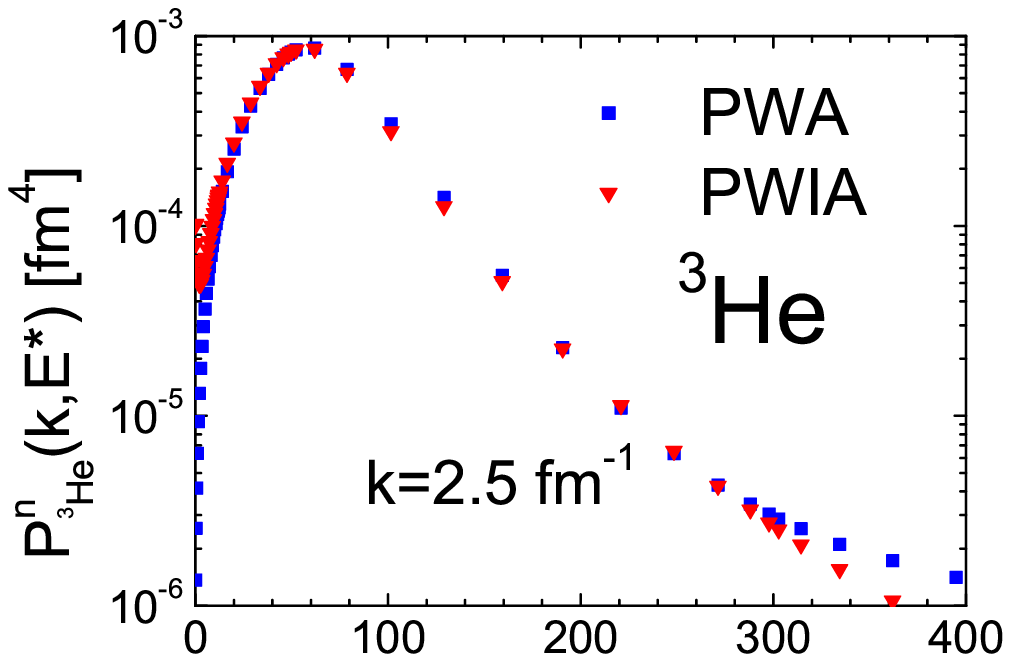}
\includegraphics[width=0.45\textwidth,keepaspectratio]{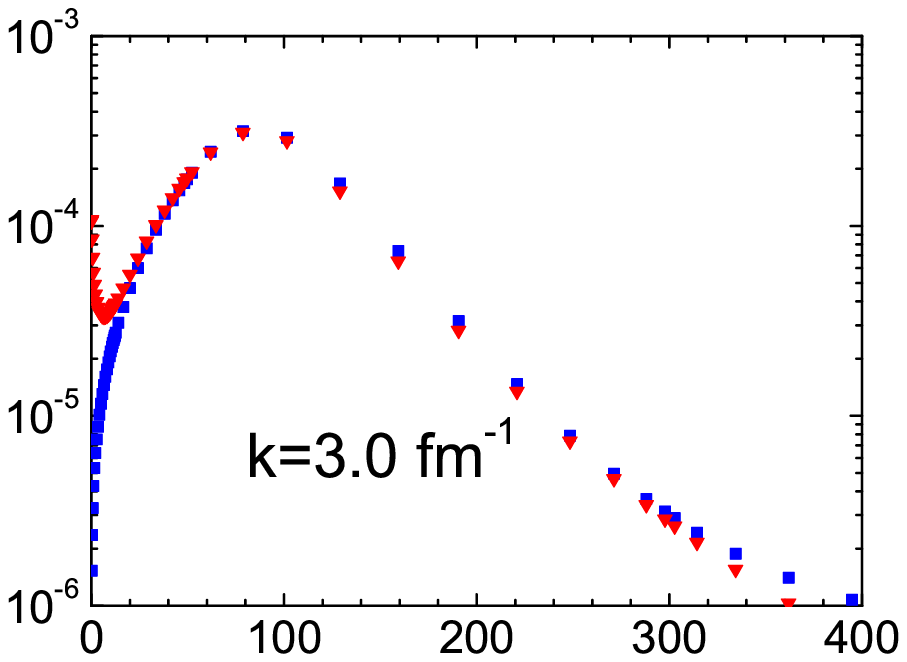}}
\centerline{\includegraphics[width=0.48\textwidth,keepaspectratio]{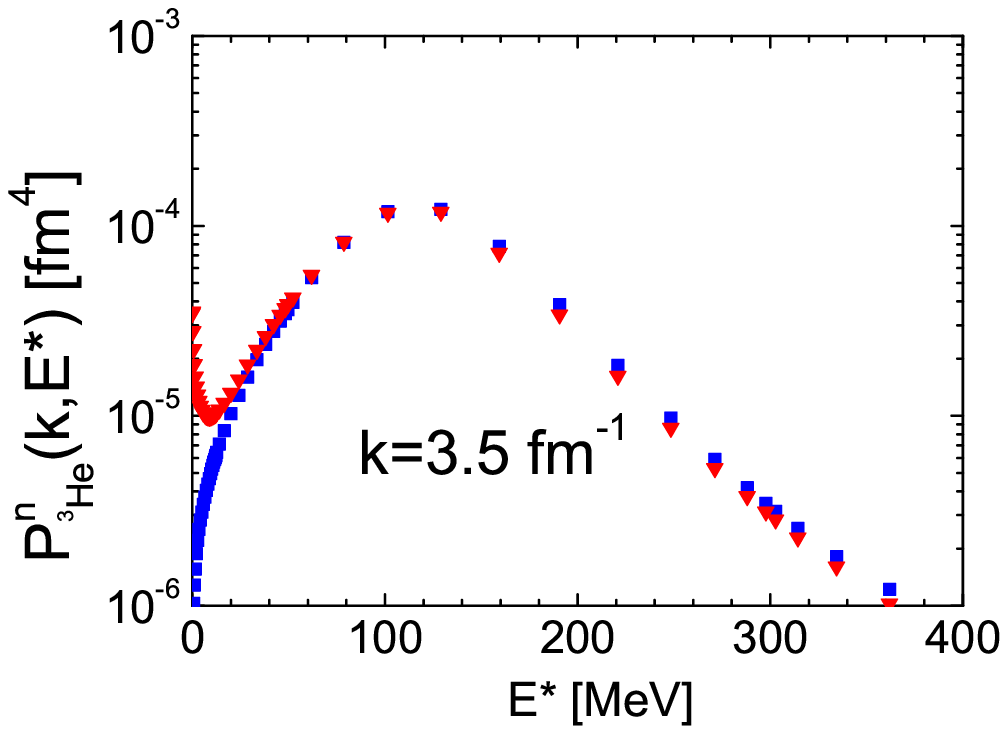}
\includegraphics[width=0.45\textwidth,keepaspectratio]{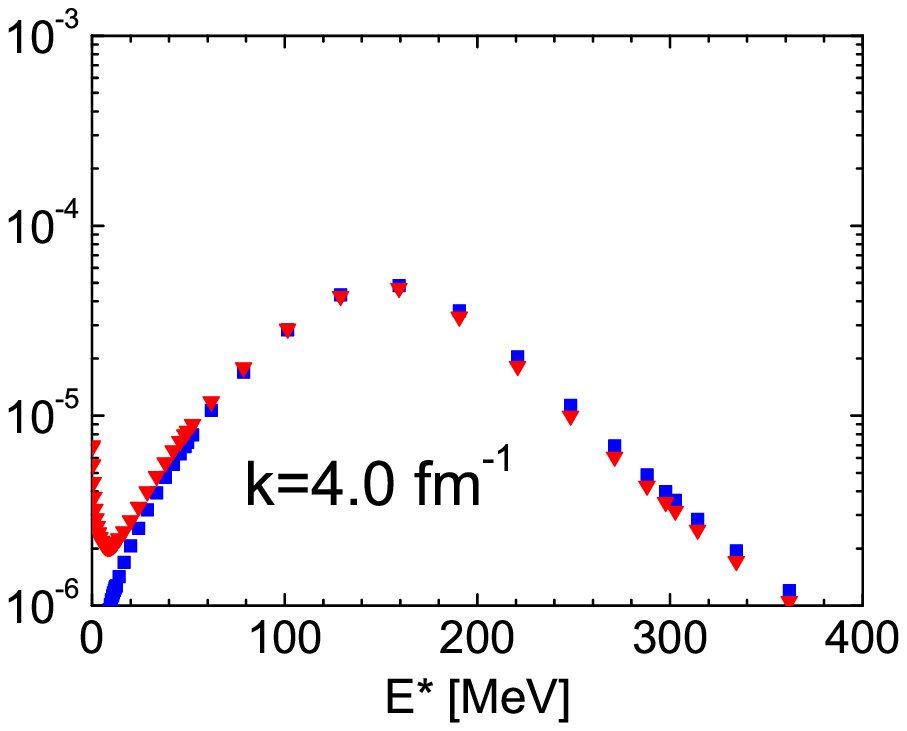}}
\vskip -0.5cm
 \caption{(Color online)
 The {\it ab initio} neutron spectral function  of
   $^3$He shown in  Fig. \ref{Fig1}(b) in correspondence of
   several  values of the neutron momentum in Plane Wave Approximation  (PWA) and
   Plane Wave Impulse Approximation (PWIA), i.e., respectively, by disregarding (blue squares) and
   including (red triangles)
   the interaction in the $pp$ final state. It can be seen that
   at high values of the neutron momentum ($k \gtrsim 2.5 \,fm^{-1}$) the final state interaction in the
    the $pp$
   essentially  affects only the   Spectral Function at small values of the  excitation energy $E^*$ (In this and
   the following Figures
    $E^* = E_{rel}$).}
\label{Fig2}
\end{center}
\end{figure}

 Due to the summation over the entire spectrum of states of the final nucleus,
 the exact ({\it ab-initio})  spectral function can only be calculated
  for the three-nucleon systems for which only two final states are open, namely the deuteron and the
  continuum two-nucleon states. For this reason  in this  paper
  we will consider the case of mirror nuclei with A=3,  which are described by
two different spectral functions and momentum distributions, namely the proton ($p$)  and the neutron ($n$) ones, which are defined as
follows
\be
 P_{3}^{p(n)}( k,E) =  P_{gr}^{p(n)}( k,E) + P_{ex}^{p(n)}( k,E),
\label{SFp}
\ee
for the proton (neutron)   spectral function in $^3$He ($^3$H), and 
\be
 P_{3}^{p(n)}(|{\b k}_1|,E) =  P_{ex}^{p(n)}(|{\b k}_1|,E),
\label{SFn}
 \ee
\noindent for the proton (neutron)  spectral function in $^3$H($^3$He). In both nuclei
 the ground ({\it gr}) part,  has the following form
\be P_{gr}^{p(n)}(|{\b k}_1|,E) = n_{gr}^{p(n)}(|{\b k}_1|)\delta(E - E_{min}),
 \label{pgr}
\ee
 where $E_{min}= |E_3|-|E_2| \approx 5.49\, MeV$ and
 $ n_{gr}^{p(n)}(|{\bf k}_1|\equiv k_1)$, is the momentum distribution corresponding  to the two-body break-up (2bbu)
 channel
$^3He \rightarrow D+p\, ({^3}H\rightarrow D+n$),  namely
\be n_{gr}^{p(n)}( k)=\frac{1}{(2 \pi)^3} \frac{1}{2}
  \sum_{M_D,M_3,\sigma_1}
  \left | \int {\rm e}^{-i\boldrho{\bk}} \chi_{\frac12 \sigma_1}^\dagger
  \Psi_{D}^{{M_D}\dagger}(\b{r} )
  \Psi_{He(H)}^{M_3}(\boldrho,\b{r})
  d \boldrho d {\bf r}\right |^2;
\label{ngr}
\ee
 here $\Psi_{He(H)}^{M_3}(\boldrho,\b{r})$ is the  $^3He$($^3H$) ground-state
  wave function,
${ M}_3$  the projection of the  spin of  $^3He$ ($^3$H),   $\b{r}$ and $\boldrho$ the Jacobi coordinates describing, respectively, the
relative motion of the spectator pair and the motion of its c.m. with respect to the active nucleon \lq\lq 1\rq\rq. The second, excited
($ex$)   part $P_{ex}^{p(n)}$ of $P_{3}(|{\b k}_1|,E)$ in Eq. (\ref{SFp})
  corresponds
to the three-body break-up (3bbu) channel $^3He(^3H)\to npp(n)$ and can be written, e.g. for the neutron spectral function in $^3$He to be
considered in this paper, as follows
 \be
 P_{ex}^{n}(|{\b k}|,E)&=& \frac{1}{(2 \pi)^3} \frac{1}{2}
  \sum_{{M}_3, S_{23},\sigma_1}
       \int \frac{d^3 \b {t}}{(2\pi)^3}
  \left | \int  {\rm e}^{-i\boldrho{\bk}} \chi_{\frac12 \sigma_1}^\dagger
 \Psi_{pp}^{\b{t}\dagger}(\b{r}) \Psi_{He}^{{M}_3}(\boldrho,\b{r})
    d\boldrho d {\bf r}\right |^2\times\nonumber\\
&\times&\delta \left( E - E_3 - \frac{\b {t}^2}{m_N}  \right),
\label{piex}
 \ee
 where $\Psi_{pp}^{\b{t}}(\b{r})$ is  the two-body
spectator continuum wave functions
 characterized by spin projection $S_{23}$ and by the relative momentum
${\b t}= \frac {{\b k}_2 - {\b k}_3}{2}$ of the $pp$ pair in the continuum.
  This  definition of the SF, used in this and in  other papers on the subject,
  is referred to as the {\it plane wave impulse
  approximation (PWIA)}, in which  the continuum wave function of the spectator pair
  in the final state has to be chosen as the  exact solution
    of the same Hamiltonian  used to obtain
  the  ground-state wave functions, with the motion of the active nucleon in the final state  described
  by a plane wave;
   if,  moreover,  the interaction in the spectator pair is  disregarded, with
     the three nucleons in the final state   described by plane waves,  one is referring
     to  the so called
   {\it plane wave
  approximation (PWA)},  a case which is relevant for the coming discussion. As a matter of fact,
  we are interested in the problem as to  whether and to which  extent the  PWIA can be approximated  by the PWA, since
  the microscopic  model of the SF  we are going to present implies the validity of the latter.  In Fig. \ref{Fig1}
  two   theoretical neutron SFs of $^3$He are shown,
 namely the one obtained with a 3N   variational wave function  \cite{CPS} corresponding to the Reid Soft Core (RSC) interaction
 \cite{Reid},
 and the  one obtained  \cite{Ciofi_Kapta} using {\it ab initio} 3N wave functions
  \cite{Kievsky:1992um} corresponding to
  the AV18 \cite{Wiringa:1994wb} NN interaction. It can be seen that,  at high values of $k$ and $E^*$, both  SFs exhibit two common features, namely : (i) a peak  located
at values of the removal  energy equal to $E \simeq k^2/4m_N$, and, more importantly,  (ii) almost identical values around the peak of the
PWIA and the PWA predictions, which means that around the peak  the two-nucleon final state can safely be approximated by plane waves. This
similarity  between the PWIA and the PWA,  which is  illustrated in more  detail in Fig. \ref{Fig2} in correspondence of several values of
the momentum,    will be shown in what follows to represent a clear manifestation of SRC.
\section{The kinematics of two- and three- nucleon SRC}

In this  Section it   will be  shown that
 different kinematical features  of 2N and 3N SRC will differently affect the
  momentum and removal
 energy distributions of  $P_{ex}^N(k,E)$.
\subsection{2N SRC}
Momentum conservation in a system of $A$ interacting nucleons
implies that
\beq \sum_{i=1}^A {\bf k}_i =0,
\label{totalKappa}
\eeq
with  the relative and c.m. momenta of a correlated  nucleon-nucleon  pair being 
 \beq
 {\bf k}_{rel}=
\frac{{\bf k}_1-{\bf k}_2}{2}  \qquad {\bf K}_{c.m.}={\bf
k}_1+{\bf k}_2= -\sum_{i=3}^A{\bf k}_i \equiv -{\bf K}_{A-2}.
\label{rel_cm_momenta}
\eeq
 It is a common practice to assume  \cite{Frankfurt:1988nt} that  2N SRC represent  those configurations, depicted in Fig.
\ref{Fig3}(a), in which the active, high momentum nucleon
 \lq \lq 1 \rq \rq  is correlated with the high momentum nucleon  \lq\lq 2 \rq\rq,  with resulting
\lq\lq high \rq\rq  relative momentum ${\bf k}_{rel}= [{\bf k}_1 -{\bf k}_2]/2$,  and \lq\lq low \rq\rq c.m. momentum ${\bf K}_{c.m.}={\bf
k}_1+{\bf k}_2=  -{\bf K}_{A-2}$. Assuming that the  $(A-2)$ nucleus is left in its ground state, the intrinsic excitation energy of the
$(A-1)$-nucleon system $E_{A-1}^*$ is given by the relative kinetic energy of the system composed by the second correlated nucleon
${\textrm N}_2$ (with momentum ${\bf k}_2$) and the $(A-2)$-nucleon system  (with momentum $\bf{K}_{A-2}$),  namely
\beq  E_{A-1}^*=\frac{1}{2\,m_N}\frac{A-2}{A-1} \left[{\bf k}_{2}-\frac {{\bf K}_{A-2}}{A-2}\right]^2\,
 \xrightarrow{A=3}  \,\frac{1}{m_N}\left(\frac{\textbf{k}_2-\textbf{k}_{3}}{2}\right)^2.
 \label{2Nkin}
 \eeq
  In the case of the so called \lq\lq naive 2NC model \rq\rq, which is the model based
  upon the assumption that
$\textbf{K}_{A-2}=0$ ($\textbf{k}_2=-\textbf{k}_1 \equiv -\textbf{k}$)\, Eq. (\ref{2Nkin}) trivially becomes 
\beq
 E_{A-1}^*=\frac{1}{2\,m_N}\frac{A-2}{A-1}\,{\bf k}_{2}^2\,  \xrightarrow{A=3}  \, \frac{{\bf k}^2}{4m_N}.
 \label{EX2NC}
 \eeq
For  the 3N systems, the main object of our investigations in the present paper,  the residual nucleus is just the third {\it spectator}
nucleon, so that the excitation energy of the ${\textrm (A-1)}$-nucleon system is exactly  the relative kinetic energy of particles \lq\lq
2\rq\rq and \lq\lq 3\rq\rq, i.e. $k^2/(4m_N)$, in agreement with the  non relativistic {\it ab initio} calculation of the spectral function
shown in Figs. \ref{Fig1} and \ref{Fig2}.

\subsection{3N SRC}
\begin{figure}[!h]
\centerline{\includegraphics[width=0.4\textwidth,keepaspectratio] {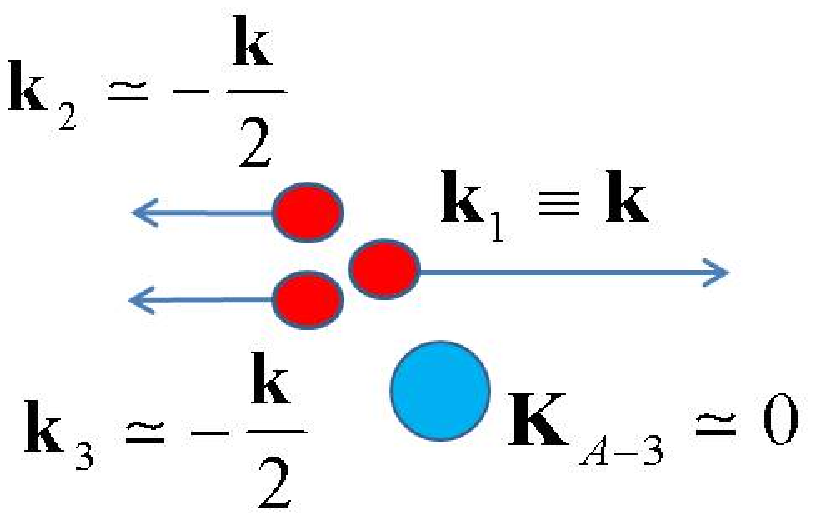}}
\begin{center}
\vskip -0.5cm ({\bf a})
\end{center}
\vskip 1.0cm
\centerline{
\includegraphics[width=0.4\textwidth,keepaspectratio] {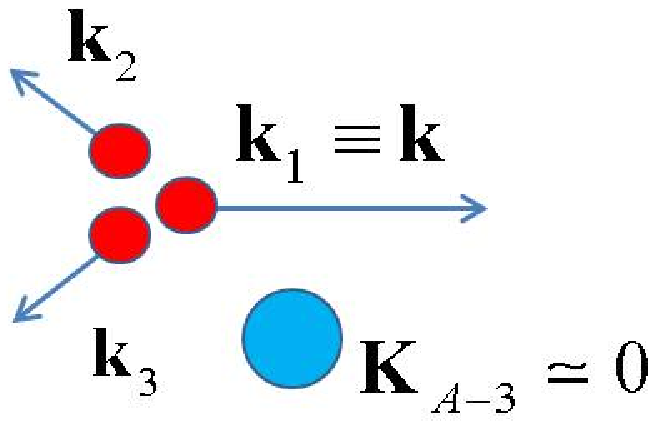}
\hskip 2.0cm
\includegraphics[width=0.4\textwidth,keepaspectratio]
{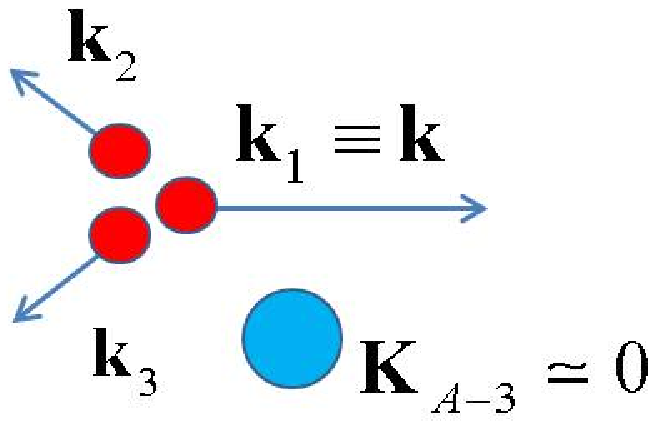}} \vskip 0.5 cm
\begin{center}
(${\bf b_1}$) \hskip 8cm (${\bf b_2}$)
\end{center}
 \vskip 1.3cm \caption{(Color online) ({\bf a}) Pictorial
representation of the kinematics of 2N SRC in nucleus A:
 the high momentum $\textbf{k}_1
\equiv \textbf{k} \gtrsim 2.0 \, fm^{-1}$ of  nucleon \lq\lq
1\rq\rq is almost completely balanced by the momentum
$\textbf{k}_2 \simeq -\textbf{k}$ of the correlated partner
nucleon \lq\lq 2 \rq\rq, with  the residual system
 moving  with low momentum $|\textbf{K}_{A-2}|
= |\textbf{k}_1+\textbf{k}_2| \lesssim 1.0 \, fm^{-1}$.
  Momentum conservation
  reads as follows: $\sum_1^A\:\textbf{k}_i=\textbf{k}_1+\textbf{k}_2+\textbf{K}_{A-2}=0$.
   In case of $A=3$  the $(A-2)$ nucleus is just a nucleon
    with momentum $\textbf{K}_{A-2} =\textbf{k}_{3}$.\\
({\bf b}) Pictorial  representation of the kinematics of 3N SRC
in
 nucleus A. The three nucleons have high momenta
and low c.m. momentum which is balanced by the momentum of the
system $A-3$. In the case of $A=3$ the  configuration in (${\bf b_1}$) can
affect only the low removal energy  part of the spectral function,
whereas in the configuration  (${\bf b_2}$) also the high removal energy part
can be affected.}
\label{Fig3}
\end{figure}

When 3N SRC are at work, two limiting cases should be considered. In the first one \cite{Frankfurt:1988nt}, depicted in Fig. \ref{Fig3}
(b$_1$),
 the high momentum ${\bf k}$
of the active nucleon is balanced by two nucleons having almost
equal momenta ${\bf k}/2$ antiparallel to ${\bf k}$; in such a
configuration the excitation energy of (A-1) is   trivially given
by
 \beq
E_{A-1}^*=\frac{A-3}{A-1}\,\frac{{\bf k}^2}{4\,m_N},
\label{EX3NC_FS}
 \eeq
  which, obviously,  vanishes for the
three-nucleon system, being zero the relative momentum of particles "2" and "3". In the second, more general case depicted in Fig.
\ref{Fig3}(b$_2$),
 the excitation energy of (A-1) will be
 \beqy
 && \hskip -0.8cm E_{A-1}^*\nonumber\\
&& \hskip -0.8cm =\frac{1}{2\,m_N}\frac{A-2}{A-1} \left[{\bf k}_{2}-\frac {{\bf K}_{A-2}}{A-2}\right]^2 + \frac{1}{2\,m_N} \frac{A-3}{A-2}
 \left[{\bf k}_{3}-\frac{1}{A-3}{{\bf K}_{A-3}}\right]^2
\xrightarrow{A=3}\,\frac{1}{m_N}\left[\frac{{\bf k}_2-{\bf k}_{3}}{2} \right]^2 .
 \label{EX3NC}
\eeqy
Thus in the  case of Fig. \ref{Fig3}(b$_2$)
 the high momentum ${\bf k}$
of the active nucleon is balanced by two nucleons with high relative  momentum $({\bf k_2}-{\bf k_3})/2$ ,  with resulting high excitation
energy of $(A-1)$ given by Eq. (\ref{EX3NC}), and
 3N SRC  are expected to  affect  the high removal energy sector
of the 3N spectral function in a way that will be  illustrated in
the next Sections.
 \section{Factorization of the many-body  wave function  in the correlations region
 and the convolution structure of the  spectral function and momentum distribution}
\subsection{Factorization: the fundamental property of
 the nuclear wave function in the correlation region}
As  previously mentioned, several  recent papers have   argued
\cite{Newpaper,CiofidegliAtti:2010xv,Baldo:1900zz,CiofidegliAtti:1995qe,Weiss:2015mba} that at short inter-nucleon
 relative distances the ground-state realistic many-body nuclear wave function $\Psi_o$ exhibits the property of
 {\it factorization},  namely
\beqy \lim_{r_{ij}\rightarrow 0}  \Psi_0(\{\Vec r\}_A) \simeq \mathcal{\hat A}\Big\{\chi_o(\Vec R_{ij})
\sum_{n,f_{A-2}}a_{o,n,f_{A-2}}\Big[ \Phi_n(\Vec x_{ij},\Vec r_{ij}) \oplus\Psi_{f_{A-2}}(\{\Vec x \}_{A-2},\{\Vec r \}_{A-2})\Big] \Big\},
\label{wf_fact}
 \eeqy
which, in turns,  is the origin of  the presence of high momentum
 components
 \cite{CiofidegliAtti:1995qe,Weiss:2015mba}.
 In Eq.
(\ref{wf_fact}):
 {i) $\{\Vec r\}_A$ and $\{\Vec r\}_{A-2}$ denote the set of radial coordinates of nuclei $A$ and $A-2$,
respectively; (ii)   $\Vec r_{ij}$ and $\Vec R_{ij}$
 are the relative and c.m. coordinate of the nucleon pair $ij$, described, respectively, by the
relative  wave function $\Phi_n$ and the c.m. wave function
$\chi_o$  in $0s$ state; iii) $\{\Vec x \}_{A-2}$ and  $\Vec
x_{ij}$ denote the set of  spin-isospin coordinates
 of the nucleus  $(A-2)$ and of the pair $(ij)$. Factorized wave functions have been introduced in the past
 as physically sound
 approximations of the unknown nuclear wave function (see e.g. \cite{Levinger}), without however providing any evidence
 of the validity  of such an approximation due to the lack, at that time, of realistic solutions of the nuclear many-body problem which,
 however became recently available and  the quantitative
 validity of the factorization approximation could be quantitatively checked. As a matter of fact
the factorization property of  realistic many-body  wave functions has been proved  to hold in the case of {\it ab initio} wave functions
of few-nucleon systems \cite{CiofidegliAtti:2010xv}  and in nuclear matter treated within the Brueckner-Bethe-Goldstone approach
\cite{Baldo:1900zz}. Moreover it has been shown \cite{Newpaper} that the 2N momentum distribution
 in light nuclei in  the region of {\it high} ($k_{rel}\gtrsim 2 fm^{-1}$) relative  momentum  obeys indeed the property of
factorization, i.e. it becomes independent upon the angle $\Theta$} between ${\bf k}_{rel}$ and ${\bf K}_{c.m.}$, namely
\beqy
n_A^{N_1N_2}({\bf k}_1,{\bf k}_2)=n_A^{N_1N_2}({ k}_{rel}, { K}_{c.m.}, \Theta) \simeq
  n_{rel}^{N_1N_2}({ k}_{rel})\, n_{c.m.}^{N_1N_2}({ K}_{c.m.})
  \label{factorization}
\eeqy
 which, in the case of $pn$ pairs, becomes
\beq
 n_A^{pn}({ k}_{rel},{ K}_{c.m.})\simeq C_A^{pn}
  n_D({ k}_{rel})\,n_{c.m.}^{pn}({K}_{c.m.})
\label{factpn}
 \eeq
 where $n_{D}$ is the deuteron momentum distribution  and $C_A^{pn}$ is a constant depending upon the
 atomic weight and which, together with the integrals of $ n_{D}({ k}_{rel})$ and  $n_{c.m.}^{pn}({ K}_{c.m.})$
 in the proper SRC region,  counts the number of SRC $pn$ pairs in the given nucleus. It should  be stressed that
 Eq. (\ref{factpn}) is  free from any adjustable
parameters  since all quantities
 appearing there
result from many-body calculations \cite{Newpaper}. It should also be stressed that the results
of Ref. \cite{Newpaper} demonstrate that
 factorization is valid in the  range of momenta including both low and high  values of the c.m. momentum; in particular, as it will be
quantified later on,  the minimum value  of the relative momentum at which
  factorization starts to occur is a function  of the value of the c.m. momentum $K_{c.m.}$, namely
  factorization is valid  when $k_{rel} \gtrsim k_{rel}^-(K_{c.m.})$, with \cite{Newpaper}
  \beqy
  k_{rel}^-(K_{c.m.}) \simeq a+b\,\phi(K_{c.m.}),
  \label{kameno}
\eeqy
where $a \simeq 2 \, fm^{-1}$ and the function $\phi(K_{c.m.})$ is such that  $\phi(0) \simeq 0$. The factorization of the  momentum
distributions leads to  an interesting and physically
 sound  interpretation,
 namely the region of ${\it high}$ relative  and ${\it low}$
 c.m. momenta is governed by
2N SRC,
 whereas  the region in which also the c.m. momenta are  ${\it high}$
 is governed by 3N SRC.
Factorization   leads to a peculiar relationship between
 the one- and two-nucleon momentum distributions in that
 the exact relation between the  two quantities
 given by \cite{CiofidegliAtti:1995qe,Weiss:2015mba} ( $N_1\neq N_2$)
\beqy
\hspace{-1cm}n_A^{N_1}({\bf k}_1)=\frac{1}{A-1}
 \left[\int n_A^{N_1N_2}({\bf k}_1,{\bf k}_2)\,d\,{\bf k}_{2}
+2\int n_A^{N_1N_1}({\bf k}_1,{\bf k}_2)\,d\,{\bf k}_{2}\right]
\label{enne1_general}
\eeqy
can be expressed
 \textit{in the factorization
region} in terms
 of the following convolution integral
 (${\bf k}_1+{\bf k}_2 +{\bf k}_3=0$, ${\bf k}_3 =
 {\bf K}_{A-2}= -{\bf K}_{c.m.}=-({\bf k}_1+{\bf k}_2$))
\cite{CiofidegliAtti:1995qe,Newpaper}
\beqy
n_A^{N_1}({\bf k}_1)&=& \left[\int
n_{rel}^{N_1N_2}(|{\bf k}_1-\frac{{\bf K}_{c.m.}}{2}|)
n_{c.m.}^{N_1N_2}({\bf
K}_{c.m.})\, d\,{\bf K}_{c.m.}\right.\nonumber\\
 &+& \left.2\int n_{rel}^{N_1N_1}(|{\bf k}_1-\frac{{\bf K}_{c.m.}}{2}|)
  n_{c.m.}^{N_1N_1}({\bf K}_{c.m.})\, d\,{\bf K}_{c.m.}\right] \equiv n_{ex}^{N_1}({\bf k}_1),
\label{ennep}
\eeqy
so that  the {\it correlation part} of the  nucleon  spectral
function will be given by the following expression \cite{CiofidegliAtti:1995qe,Newpaper}
\beqy P_{1} ^{N_1}({\bf k}_1,E)&=&\sum_{N_2=p,n}\,C_{N_1N_2} \int
n_{rel}^{N_1N_2}(|{\bf k}_1-\frac{{\bf K}_{c.m.}}{2}|)
  n_{c.m.}^{N_1N_2}({\bf K}_{c.m.})d\,{\bf K}_{c.m.}\,\nonumber\\
  &\times&\delta
\left( E-E_{thr}- \frac{A-2}{2m_N(A-1)} \left[{\bf k}_1-\frac{(A-1){\bf K}_{c.m.}}{A-2} \right]^2\, \right)
 \label{SF_CONV}
\eeqy
where $C_{N_1=N_2}=2$ and $C_{N_1\neq N_2}=1$. This is the {\it convolution model} of the spectral function which has been first obtained
in Ref. \cite{CiofidegliAtti:1995qe} and  applied there within the following approximations: (i) an effective two-nucleon momentum
distribution for both $pn$ and $pp$ pairs has been used, and  (ii) the constraint resulting from Eq. (\ref{restriction}) has not been
considered. The limits of validity of these approximations will be discussed in what follows and in a forthcoming paper devoted to complex
nuclei.  Moreover, up to now the   convolution formula (\ref{SF_CONV}) has been applied assuming for the c.m. distribution a soft behavior
in order to enhance the effects of 2N
 SRC involving low c.m. momentum components,
which provides the largest contribution to the SRC peaks of the spectral function and to the high momentum part of the momentum
distributions.
 In the present paper, following the finding of Ref. \cite{Newpaper}, demonstrating that
 factorization may also occurs
at high values of the c.m. momentum  ({\it cf.} Fig.6 of Ref. \cite{Newpaper}),  we extend the factorization property to the treatment  of
3N SRC and include in the convolution formula both the  {\it soft} and the {\it hard} components of the c.m. momentum distributions,  both
resulting from {\it ab-initio} many-body calculations, as illustrated in the next Section in the case of the three-nucleon system.
\section{2N and 3N SRC in the spectral function of $^3$H${\bf e}$}
In this Section the  microscopic convolution model of the spectral function of the three-nucleon system
embodying 2N and 3N SRC
 will be presented and compared with the {\it ab initio} Spectral Function. For ease of presentation
  we will discuss  the neutron (proton) spectral function of $^3$He ($^3$H), which requires  only  the knowledge
  of  the $pn$  relative and c.m. momentum distributions.
\subsection{The microscopic neutron spectral function  of $^3$He within the convolution model embodying  2N and 3N SRC}
\begin{figure}
\begin{center}
\includegraphics[width=1.0\textwidth,keepaspectratio] {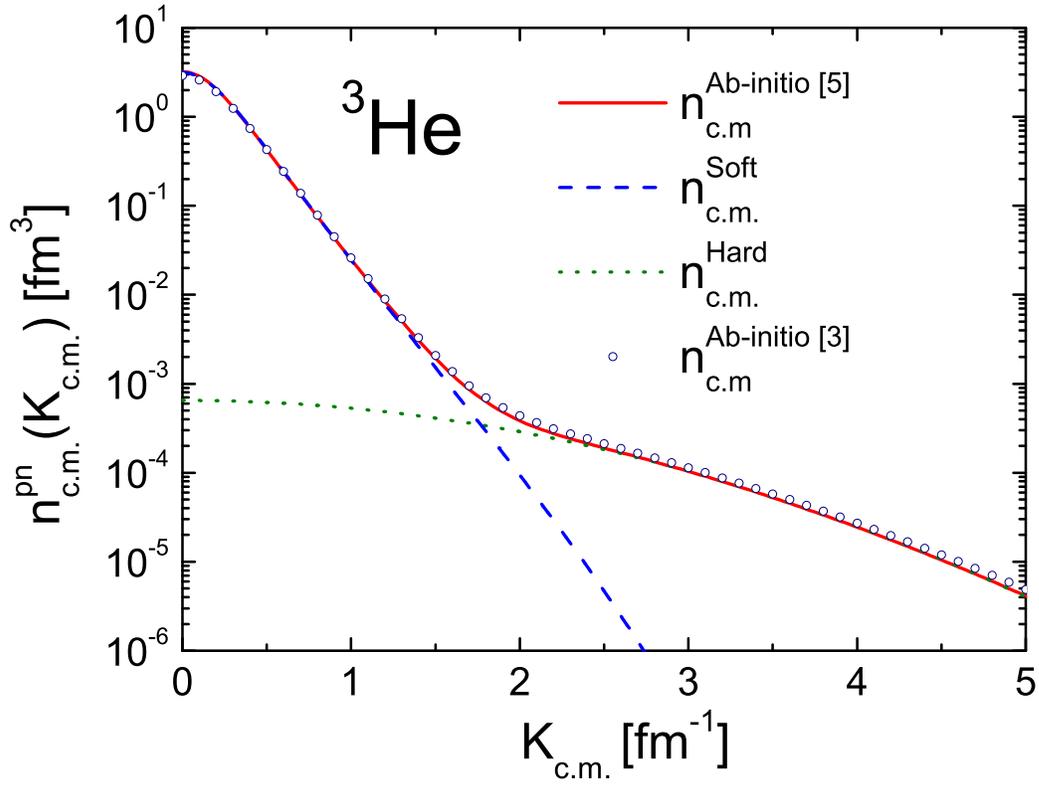}
\vskip -0.3cm \caption{(Color online) The c.m. momentum distribution of the correlated proton-neutron  pair in $^3$He calculated in Ref.
\cite{Newpaper}(full line) and in Ref. \cite{Wiringa:2013ala} (open dots)  with {\it ab initio} wave functions corresponding to the AV18
interaction. The figure shows the separation into the Soft (dashed line) and Hard (dotted line) components. (Adapted from
Ref.\cite{Newpaper})}.
 \label{Fig4}
  \end{center}
\end{figure}

\begin{figure}
\centerline{
\includegraphics[width=0.5\textwidth,keepaspectratio]
{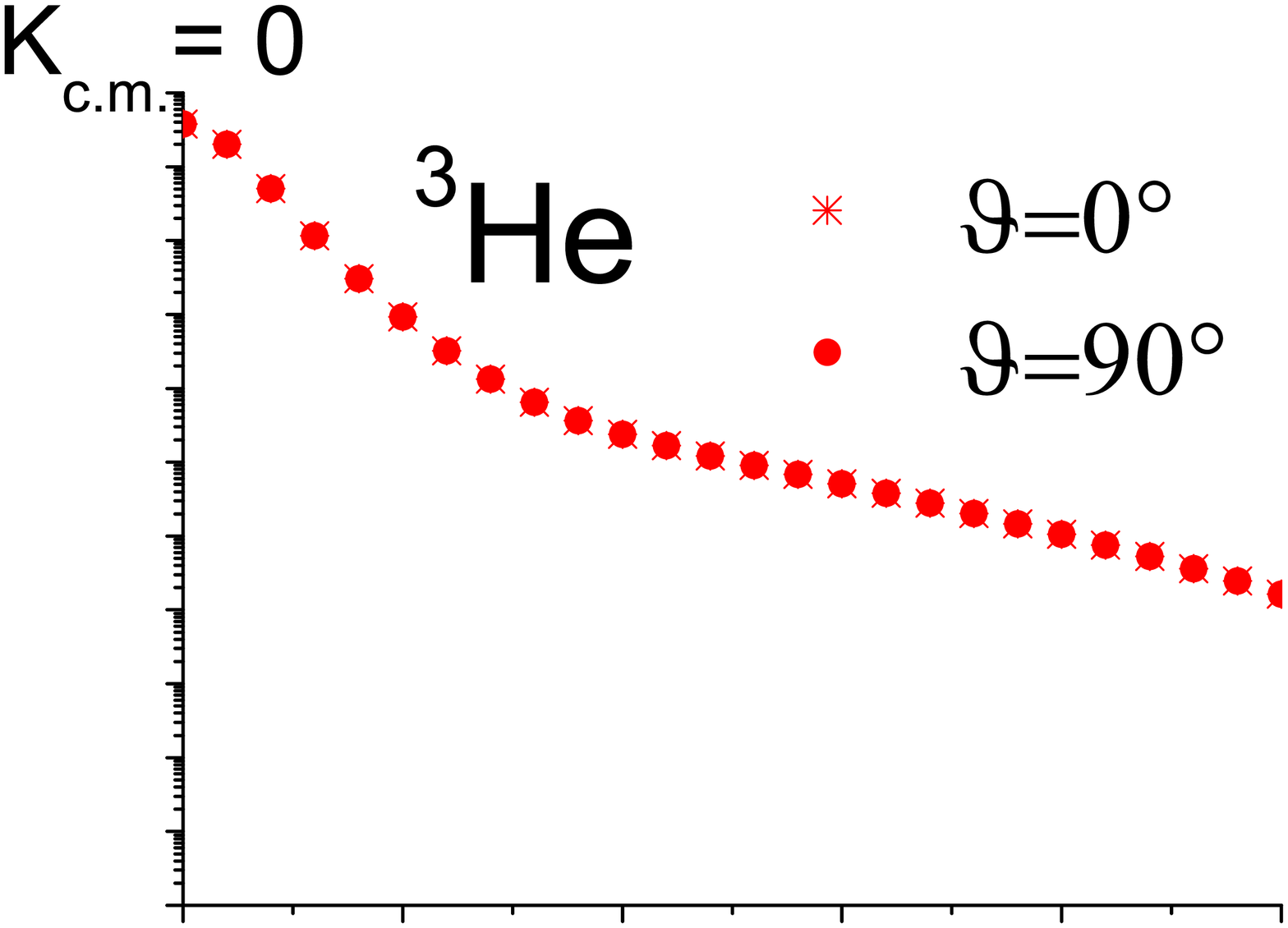}}
 \vskip -5.0cm \hskip -2.0cm
 \includegraphics[width=0.5\textwidth,keepaspectratio]
 {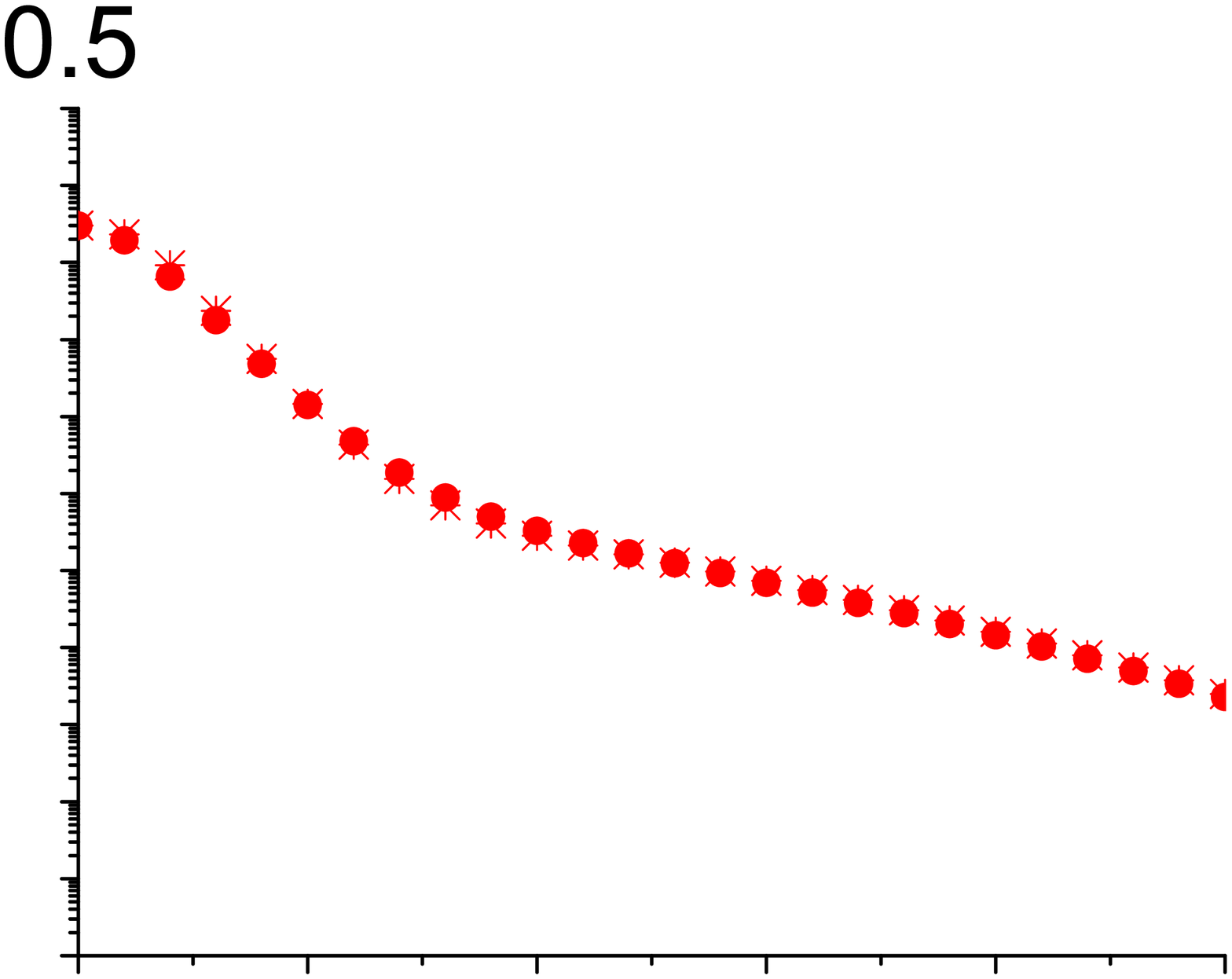}
 \vskip -5.0cm \hskip -4.0cm
 \includegraphics[width=0.5\textwidth,keepaspectratio]
 {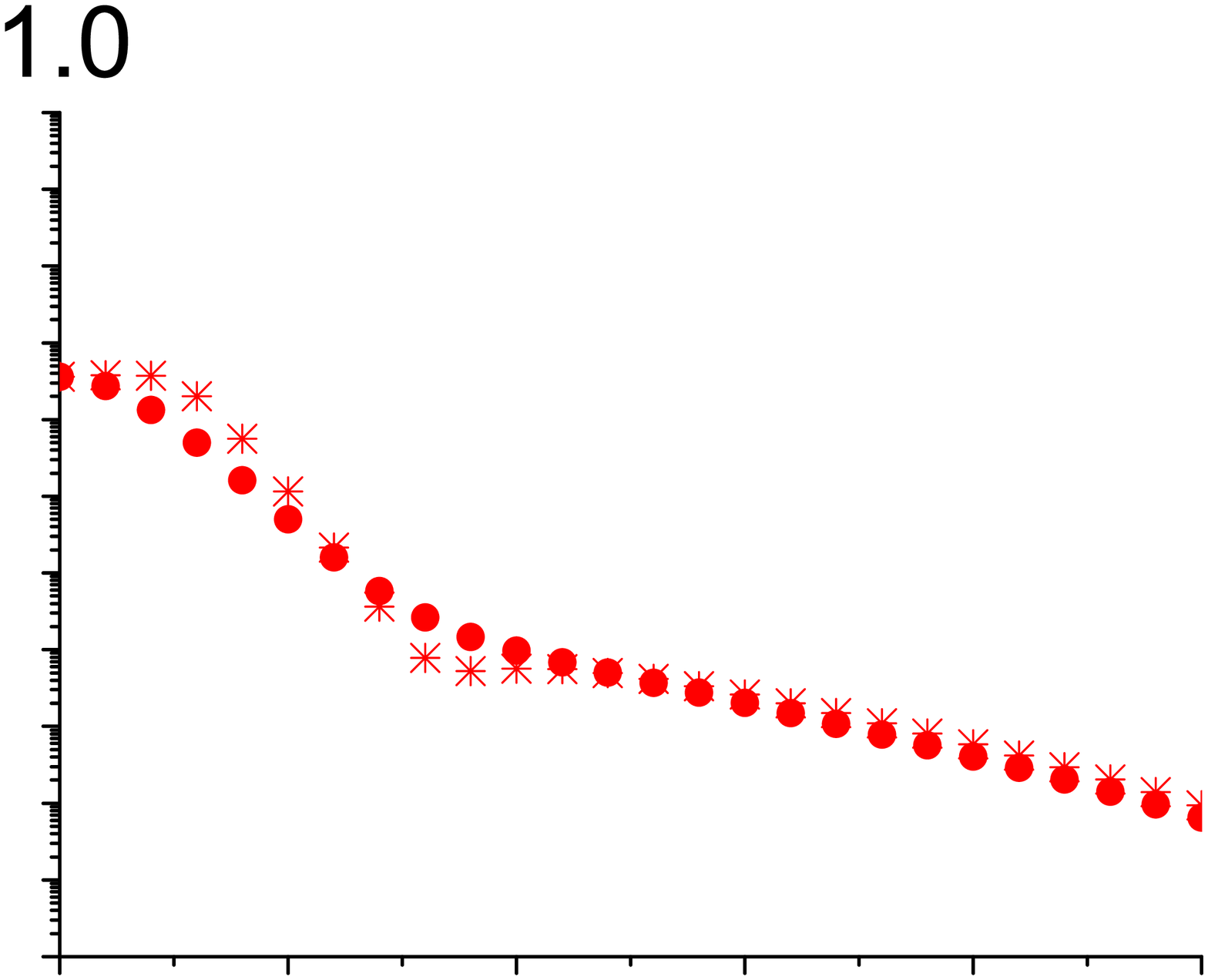}
 \vskip -5.0cm \hskip -6.0cm
 \includegraphics[width=0.5\textwidth,keepaspectratio]
 {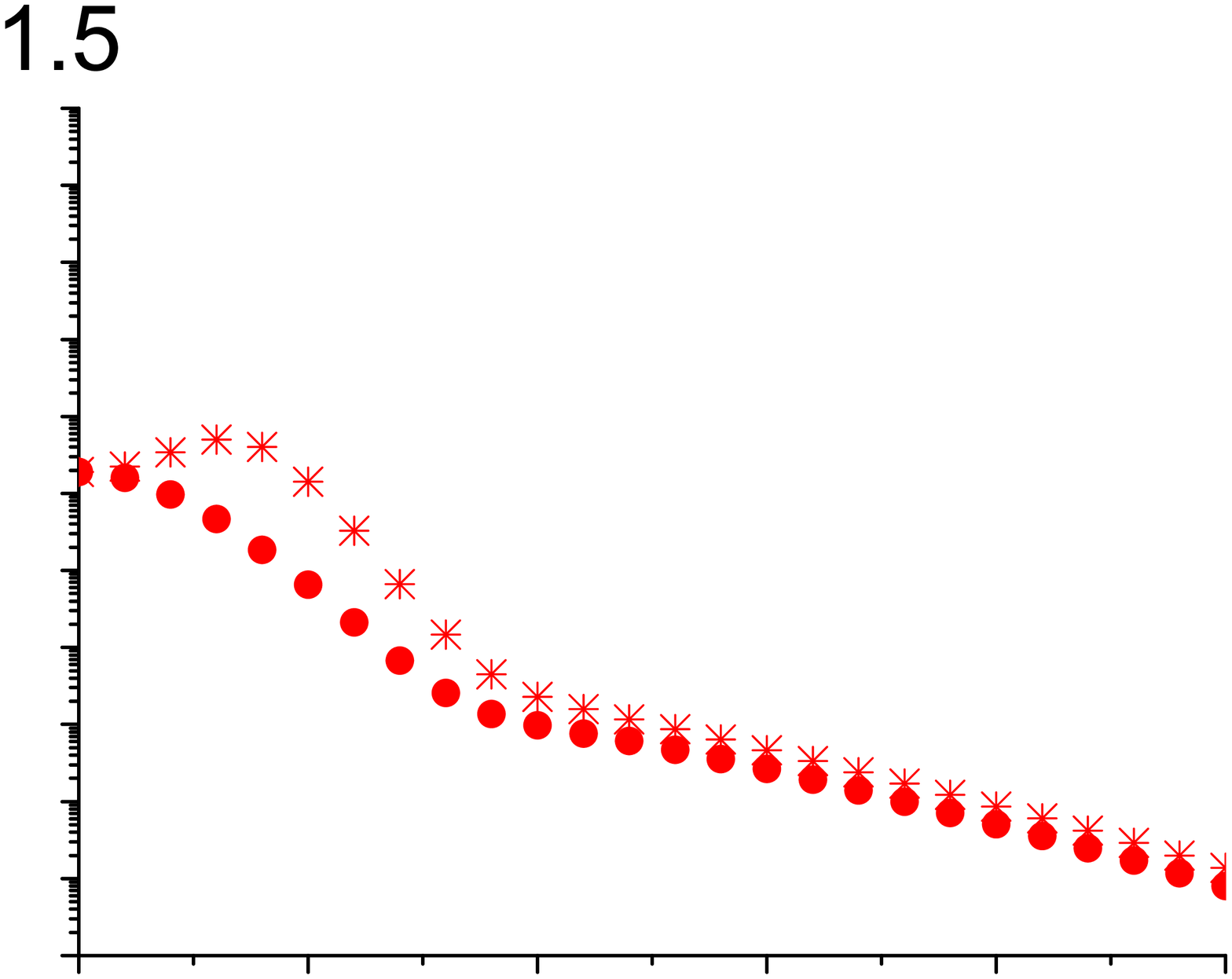}
 \vskip -5.0cm \hskip -8.0cm
 \includegraphics[width=0.5\textwidth,keepaspectratio]
 {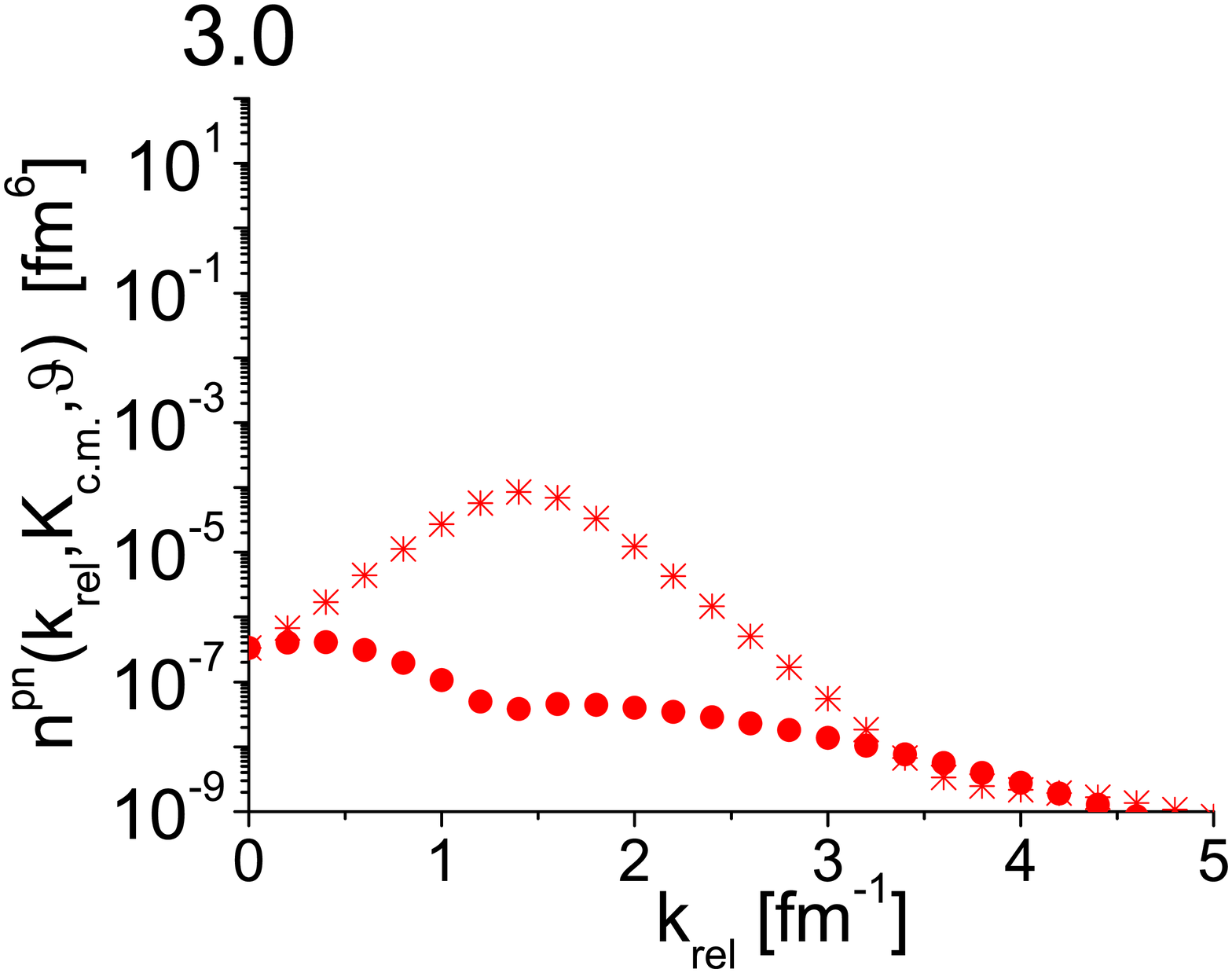}
 \caption{ (Color online) The $pn$  two-nucleon
momentum distributions  in $^3$He, $n^{pn}(k_{rel}, K_{c.m.},
\theta)$, obtained {\it ab-initio} in Ref. \cite{Newpaper} in
correspondence of several values of $K_{c.m.}$ and  two values of
the angle $\theta$ between ${\bf K}_{c.m.}$ and  ${\bf k}_{rel}$.
The region of $k_{rel}$ where the value of $n^{pn}(k_{rel},
K_{c.m.}, \theta)$ is independent of the angle determines the
region of factorization of the momentum distributions, i.e.
$n^{pn}(k_{rel}, K_{c.m.}, \theta) \rightarrow
n_{rel}^{pn}(k_{rel})n^{pn}_{c.m.}(K_{c.m.})$. It can be seen that
the region of factorization starts at  values of $k_{rel}=
k_{rel}^{-}$, which increase with increasing values of $K_{c.m.}$,
i.e.  $k_{rel}^{-} = k_{rel}^{-}(K_{c.m.})$; because of the
dependence of $k_{rel}^{-}$ upon $K_{c.m.}$, a constraint on the
region of
 integration over $K_{c.m.}$  arises from   Eq. (\ref{restriction})(Adapted from Ref. \cite{Newpaper}).}
\label{Fig5}
\end{figure}

The basic ingredients to calculate the neutron  spectral function in $^3He$ within the convolution model are the two-nucleon relative and
c.m momentum distribution of the $pn$ pair. Both quantities have been obtained  in  Ref. \cite{Wiringa:1994wb} and \cite{Alvioli:2011aa};
Fig. \ref{Fig4}  shows the  c.m. momentum distribution and it can be seen that the
 distribution can be split into a hard and a soft parts according to
\beqy n_{c.m.}^{pn}(K_{c.m.})= n_{c.m.}^{pn,soft}(K_{c.m.}) + n_{c.m.}^{pn,hard}(K_{c.m.}). \label{Total} \eeqy 
 Thus, placing Eq. (\ref{Total})  in  Eq. (\ref{SF_CONV})  the fully correlated
  neutron
 SF in  $^3$He (proton spectral function in $^3$H ) acquires the following form
 \footnote{Because of the lack of a bound $nn$ state  the sum in Eq. (\ref{SF_CONV}) extend only to free protons.}
 \beqy
   P_{ex}^{n}({\bf k}_1,E)&=& \int
n_{rel}^{np}(|{\bf k}_1-\frac{{\bf K}_{c.m.}}{2}|)
  \left[n_{c.m.}^{np,soft}({K}_{c.m.})+n_{c.m.}
  ^{np,hard}({K}_{c.m.})\right]d\,{\bf K}_{c.m.}\,\nonumber\\
  &\times&\delta
\left( E-E_{thr}- \frac{1}{4m_N} \left[{\bf k}_1-{2\,{\bf
K}_{c.m.}} \right]^2 \right).
\label{SF_CONV3}
\eeqy
Here we would like  to reiterate  that   Eq.
(\ref{SF_CONV3})
 represents a   genuine  parameter-free many-body
quantity generated by {\it ab-initio} relative and c.m. two-nucleon momentum distributions corresponding to a  given  local NN interaction.
As a matter of fact it should be remembered that the 2N relative  and c.m. momentum distributions appearing there
 are nothing but the quantities obtained
by using  the one- and two-body many-body density matrices calculated with {\it ab initio} many-body wave functions. It is also worth
stressing that Eqs. (\ref{SF_CONV}) and (\ref{SF_CONV3})
 are based upon the factorization property of the 3N wave function at short range,
  leading to the
 convolution model of the spectral function; for such a reason those equations
  are only valid in well defined
  ranges
 of the relative and c.m. momenta, which,  in the present paper, are usually quantified as follows:
 the region in which $k_{rel}^- \geq 2\,fm^{-1}$ and $K_{c.m.}
 \lesssim 1\,fm^{-1}$
  represents  the {\it 2N SRC region},  whereas   the region
 where
 $k_{rel}^- \geq 2\,fm^{-1}$ and   $ K_{c.m.}> 1\,fm^{-1}$
 identifies the 3N SRC region.
In Fig. \ref{Fig5} the $k_{rel}$ dependence of  the $pn$  momentum distributions is
 shown in correspondence of several
values of $K_{c.m.}$
 and the region of factorization satisfying the relation
  \beqy
k_{rel}\geq k_{rel}^{-}(K_{c.m.}),
\label{fac_reg}
  \eeqy

can be clearly identified as the region where the 2N momentum distributions corresponding to   $\Theta =0^o$ and $\Theta=90^o$ overlap.
Since the value of $k_{rel}^- $ depends upon the value of $K_{c.m.}$,  Eq. (\ref{fac_reg}),  generates  a constraint on the region of
integration over ${\bf K}_{c.m.}$
 in Eq. (\ref{SF_CONV3}), in that only those values of ${\bf K}_{c.m.}$ satisfying Eq. (\ref{fac_reg})
 have to be considered. Since for a fixed value of $k_1$ the relation between $k_1$ and $K_{c.m.}$ is given by
\beqy
 k_{rel} =|{\bf
k}_1-\frac{{\bf K}_{c.m.}}{2}| \geq k_{rel}^{-}(K_{c.m.}),
 \label{restriction}
 \eeqy
this  is the equation  which establishes a constraint on the
  the region of integration over ${\bf K}_{c.m.}$; this region becomes  narrower
 than the region which is obtained  if  the constraint given by Eq. (\ref{restriction}) is disregarded.
 It is worth stressing  that Eq. (\ref{restriction}) and the resulting constraint were never been considered in the past.

\subsection{The microscopic convolution model of the spectral function
of $^3$He embodying 2N and 3N SRC and its comparison with  {\it ab initio}  spectral functions}
\begin{figure}
\centerline{
\includegraphics[width=0.5\textwidth,keepaspectratio] {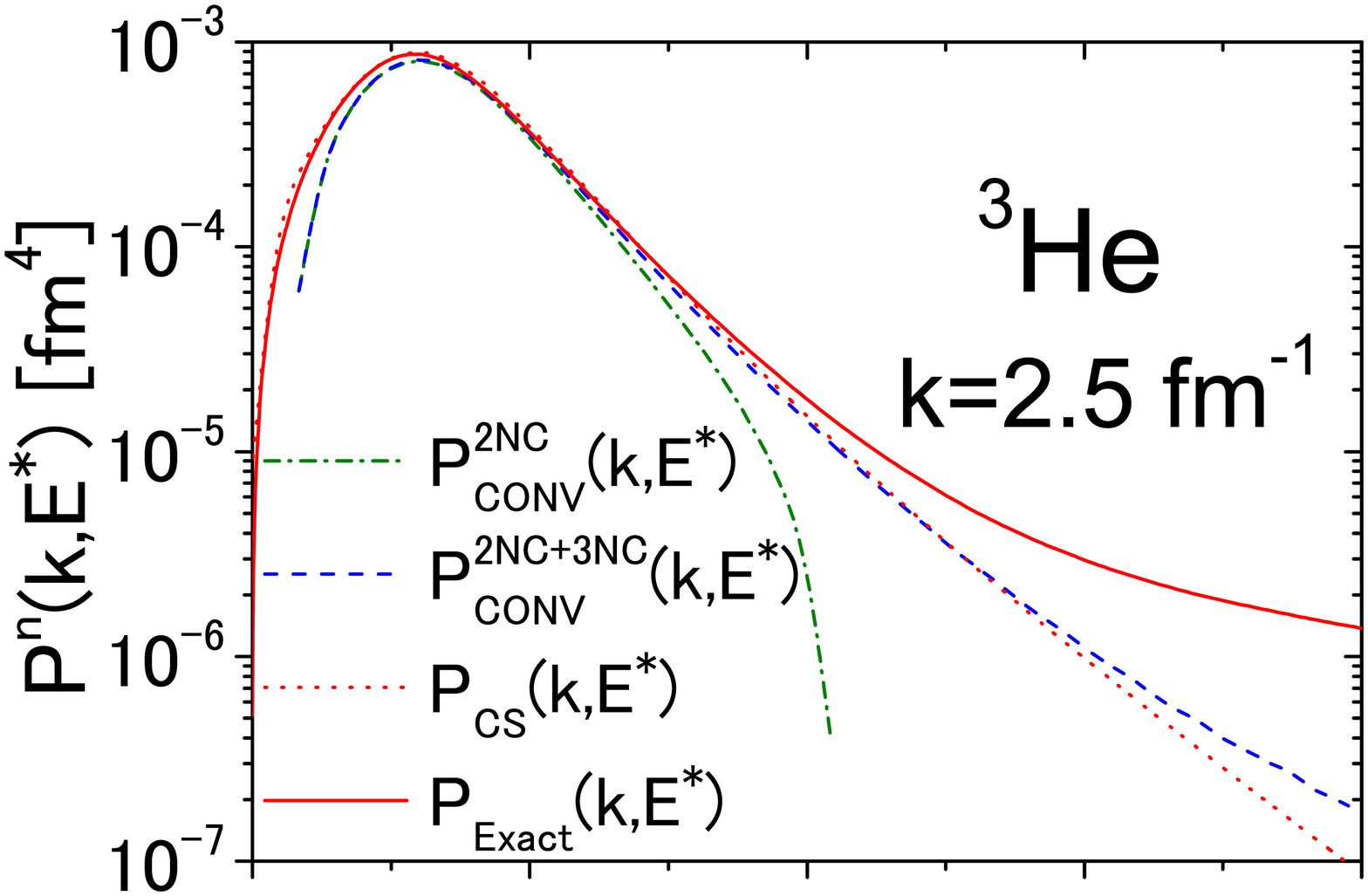}
\includegraphics[width=0.5\textwidth,keepaspectratio] {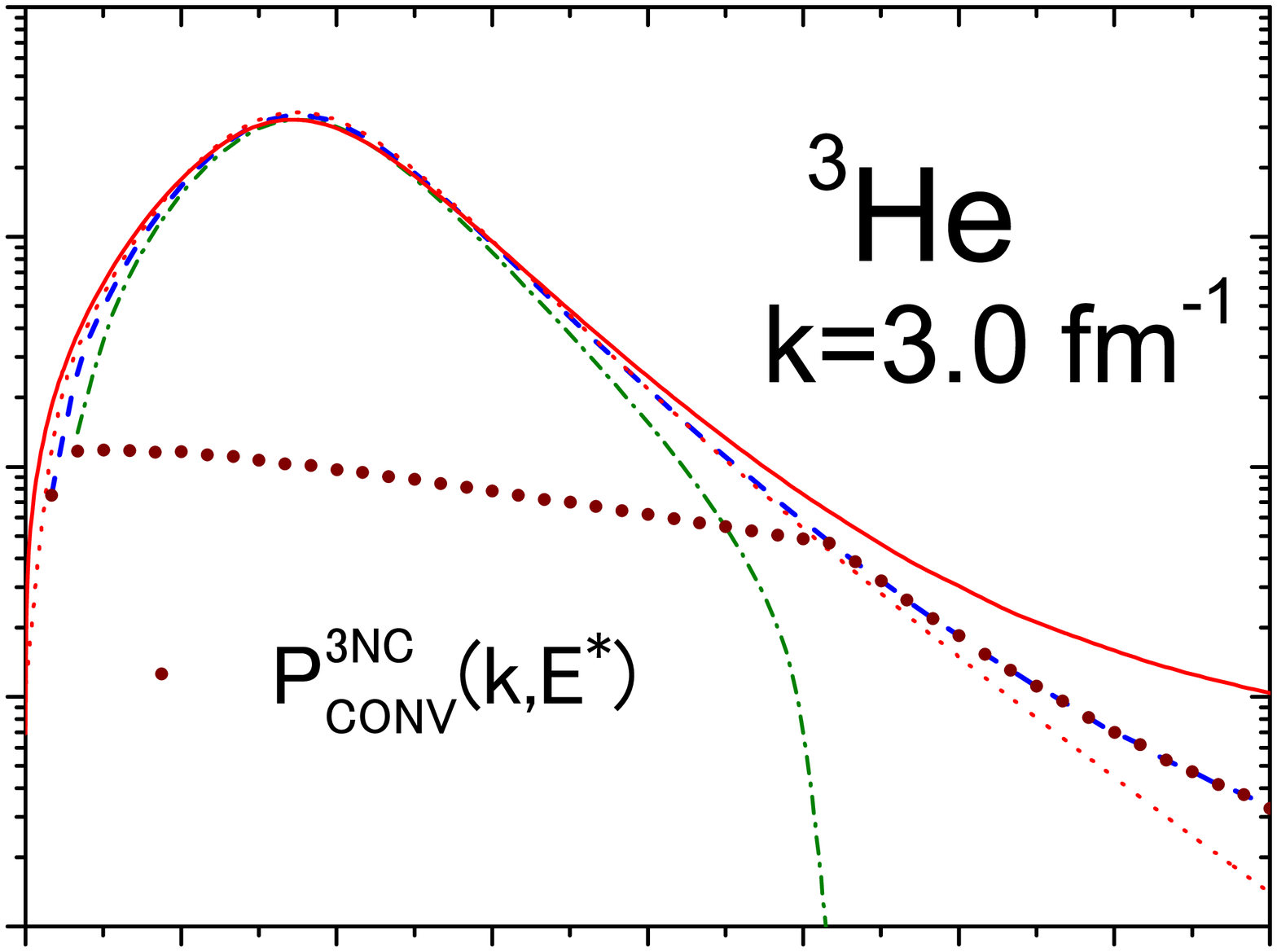}}
\vskip -1.0cm 
\centerline{\includegraphics[width=0.5\textwidth,keepaspectratio]
{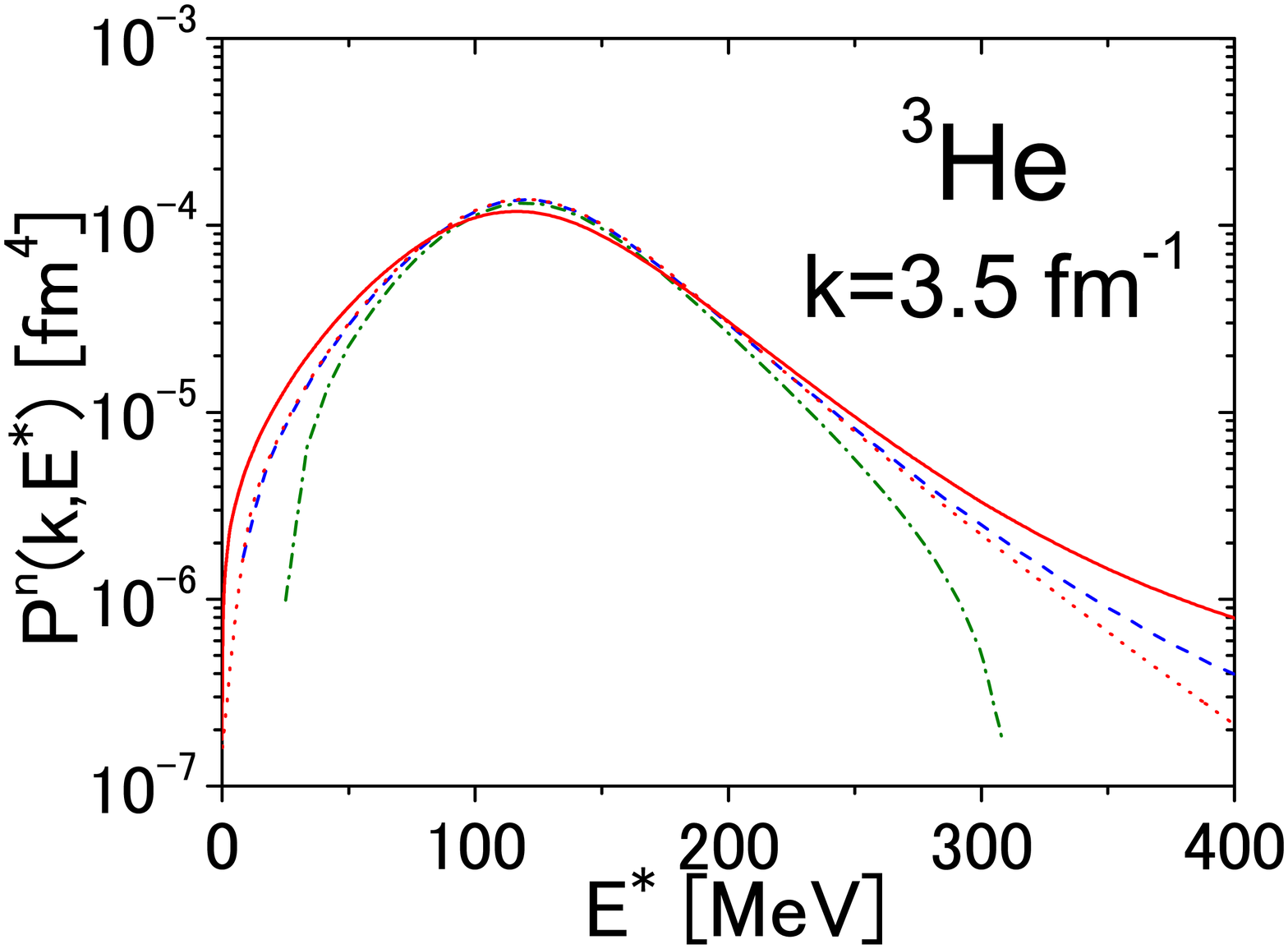}
\includegraphics[width=0.5\textwidth,keepaspectratio] {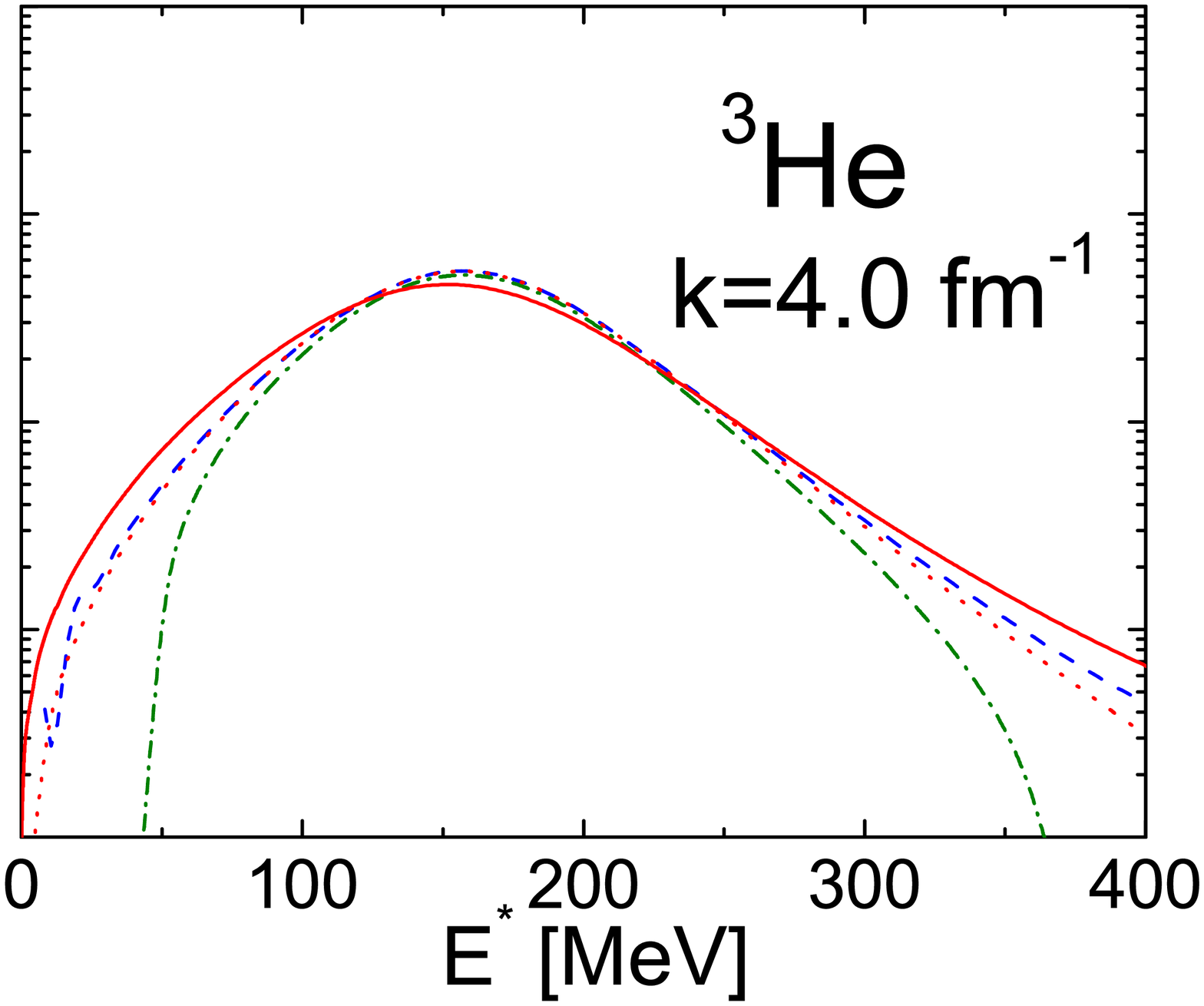}}
\vskip -0.3cm \caption{(Color online) Full line:  the {\it ab initio} spectral function  of the neutron in $^3$He in PWA corresponding to
the AV18 interaction, shown in Fig. \ref{Fig2} by the full squares. Dot-dashed line: convolution model which includes 2N SRC only ($k_{rel}
>2 \, fm^{-1}$, $K_{c.m.} \leq 1.0 \, fm^{-1}$); Dashed line: convolution model which includes  both 2N ($k_{rel} >2 \, fm^{-1}$, $K_{c.m.}
\leq 1.0\, fm^{-1})$ and 3N ($k_{rel} >2 \, fm^{-1}$, $K_{c.m.} > 1.0 \, fm^{-1})$ SRC. Both dashed and dot-dashed lines include
 the constraint on the values of ${K}_{c.m.}$
imposed by the requirement of factorization (Eq. (\ref{restriction})). Dotted line: convolution model of Ref. \cite{CiofidegliAtti:1995qe}
which uses only the soft part of the c.m. momentum distribution  without the constraint on the value of ${K}_{c.m.}$ (Eq.
(\ref{restriction})). The full dots in the case of $k=3.0 fm^{-1}$ denote the contribution from 3N SRC, i.e. the difference between the
dashed and the dot-dashed curves.} \label{Fig6}
\end{figure}

\begin{figure}
\begin{center}
\includegraphics[width=0.9\textwidth,keepaspectratio] {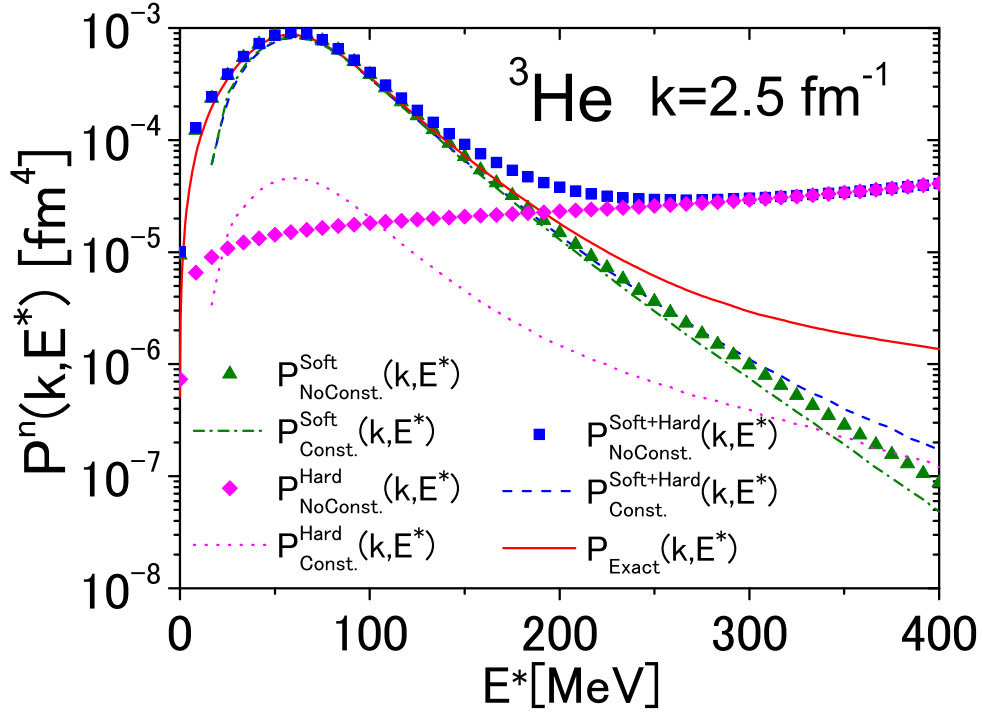}
\vskip -0.3cm \caption{ (Color online) The contributions to the
neutron spectral function of the Soft and Hard parts of the c.m.
momentum distributions shown in Fig. \ref{Fig4} in the case of
$k=2.5 \, fm^{-1}$ (cf. Fig. \ref{Fig6}) considering (Const.) and
disregarding (NoConst.) the constraint on the value of $K_{c.m.}$
generated by Eq. (\ref{restriction}).} \label{Fig6A}
\end{center}
\end{figure}

\begin{figure}
\begin{center}
\includegraphics[width=0.9\textwidth,keepaspectratio] {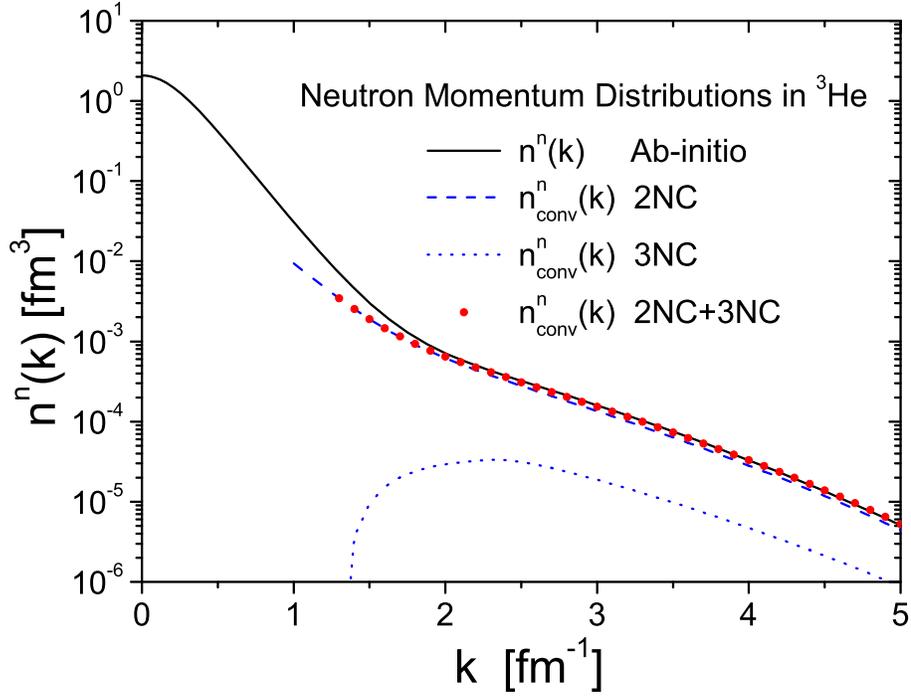}
\vskip -0.3cm \caption{ (Color online) The {\it ab-initio} neutron momentum distribution in  $^3$He \cite{Alvioli:2011aa} (full line)
compared  in the high momentum region with the distribution obtained from the momentum sum rule  (Eq. (\ref{eq6-3a})), i.e. by integrating
the convolution model spectral function (Eq. (\ref{SF_CONV})). Dotted line: contribution from 3N SRC; Dashed line: contribution from 2N
SRC; Full dots: the sum of   2N and 3N SRC contributions.} \label{Fig7}
\end{center}
\end{figure}

 In this
Section  the {\it ab initio} neutron spectral function of $^3$He,
 \cite{Ciofi_Kapta}, will be compared with the microscopic  convolution model
embodying 2N and 3N SRC calculated by Eq. (\ref{SF_CONV3}) taking properly into account  the constraint on the value of $K_{c.m.}$
 imposed by
Eq. (\ref{restriction}), unlike what done in Ref. \cite{CiofidegliAtti:1995qe}  where the constraint was not considered because the
two-nucleon momentum distribution calculated at different angles was not known at that time.
 The result of these comparisons,  in the region
 $2.5 <k < 4 \, fm^{-1}$, $E \leq 400\, MeV$,  are
  shown in  Fig. \ref{Fig6}. A careful inspection at these results suggests  the following comments:

 \begin{enumerate}
\item  the prediction by the microscopic convolution model of the spectral function   which
correctly includes 2N SRC
 as previously defined ($k_{rel} \geq 2 fm^{-1}$, $K_{c.m.} \leq 1 fm^{-1}$), as well as  the constraint resulting from Eq.
(\ref{restriction})
  (dot-dashed line in Fig. \ref{Fig6}), generally
  agrees with the  {\it ab-initio} SF (full line), as far as the  energy position of the peak, its  amplitude
  and the energy region around it are concerned, but
  severely underestimates the   high removal energy wings;

\item the inclusion of 3N SRC,  as previously defined  ($k_{rel} \geq 2 fm^{-1}$, $K_{c.m.} > 1 fm^{-1}$), into the
 convolution model  which satisfies Eq.(\ref{restriction})
(dashed line) appreciably increases the amplitudes of the wings,
leading to a  satisfactory agreement with the {\it ab-initio}
spectral function,  in a wide range of energy;
the difference between the dashed and
dot-dashed curves provides the effect of 3N SRC, whereas the
difference between  the full  and  the dashed curves identifies
the region where the 3N configurations cannot be described by the
factorized momentum distribution leading to the convolution model.

\item the  results  within the model of Ref. \cite{CiofidegliAtti:1995qe} (dotted line), where only the
soft part  of the c.m. distribution  is considered and  the constraint on the values of $K_{c.m.}$  is disregarded, do not appreciably
differ from the results obtained with the {\it ab-initio} spectral function;

\item the  results shown in Fig. \ref{Fig6} can be explained as follows: (i) the hard part  of the c.m. momentum distribution
(dotted line in Fig. \ref{Fig4}) produces  a very high and unrealistic contribution to the spectral function (see Fig.\ref{Fig6A}) which is
however cut down when the constraint (Eq. (\ref{restriction})) is taken into account; as a result, the amount of 3N SRC  produced by the
hard part of the c.m momentum distribution becomes comparable to the ones produced by the high momentum part of soft c.m momentum
distribution; (ii) the sharp decrease of the 2N SRC contribution with increasing values of E* is due to the fact that  once the integration
over the angle between $\bf{k}$ and $\bf{K_{c.m.}}$ is carried out, the limits of integrations in $K_{c.m.}$  in Eq. (\ref{SF_CONV3}) are
$K_{c.m.}^- =|k-K_0|/2$ and $K_{c.m.}^+ =(k+K_{0})/2$, with  $K_0=(4\,m_N\,E^*)^{1/2}$ and it can be trivially seen that beyond a certain
value of E*, which increases with increasing values of $k$, 2N SRC cannot occur, since they would fall outside the lower limits of
integration;

 \item  in the light of the previous remarks, it
appears that in the case of $^3$He the model of ref \cite{CiofidegliAtti:1995qe} effectively takes into account the factorization property
of the two-body momentum distributions;

\item In Fig. \ref{Fig7}  the {\it ab initio} neutron momentum distribution $n_3^n(k)$ in  $^3$He (full line) is  compared
 in the high momentum region  with
 the  distribution obtained from the momentum sum rule (Eq. (\ref{eq6-3a})), i.e.  by integrating  the
 microscopic convolution model spectral function presented
 in Fig. \ref{Fig6};   the dashed line includes only 2N SRC, whereas  the full dots include both 2N and  3N SRC;
 it can be seen that  although the contribution from 2N  SRC is almost one order of magnitude higher than  the one
 due to 3N SRC,
 the introduction of the latter brings the result of the microscopic convolution model in perfect agreement with the {\it ab-initio}
 results.
 \end{enumerate}

\section{Summary and conclusion}
The main aspects and    results of the present  paper  can be summarized as follows:
\begin{enumerate}

\item we have  reiterated  that the basis of
any treatment of SRC  is the wave function factorization at short range leading in a natural way to the  convolution model of the spectral
function and, accordingly have developed an advanced microscopic many-body, parameter-free approach to the the nucleon spectral function
 expressed in terms of {\it ab-initio} A-dependent two-nucleon relative and c.m. momentum distributions reflecting
the underlying NN interaction; by this way we take into account the specific features of the given nucleus without recurring to
approximations for finite nuclei relying on infinite nuclear matter;

\item unlike previous convolution  models of the spectral functions, in our approach the region
of factorization of the
nuclear wave function in momentum space has been clearly   identified
and the resulting constraints on the values of the relative and c.m momenta have been
properly taken into account in the convolution integral;

\item in the case of the three-nucleon system, we have found that when only 2N SRC are taken into account,
the convolution model predictions agree within 80-90 \%
 with the results of  the {\it ab-initio} spectral function as far as the peak position and the energy region around it are concerned,
 whereas far from the peak, particularly at high values
 of the removal energy, they disagree  by orders of magnitude;  this disagreement however  is strongly reduced
 when one considers
  the effects
 of 3N SRC, which are   implicitly generated by the high momentum part ($K_{c.m.} > 1 fm^{-1}$) of the soft c.m. distribution
 used  in the model of Ref.
 \cite{CiofidegliAtti:1995qe}, or arise explicitly from the introduction of the hard components of the c.m distribution
 as in Eq. (\ref{SF_CONV3}) of the present paper; it turns out that the requirement of factorization lead  to similar results in both
 cases;  and  whether such a  result remains  valid also in the case of  complex nuclei is a  current matter of investigations;

 \item we found that the high momentum part $(k\gtrsim 2\,fm^{-1})$ of the neutron momentum distribution   in $^3$He is practically
 governed by the effects of 2N SRC, since the tails of the spectral function affected by 3N SRC have small
 effects on the energy removal integration; it should however be pointed out that
 the inclusion of 3N SRC brings the result of the microscopic
 convolution model in perfect agreement with the {\it ab-initio} momentum distribution corresponding to the
 AV18 NN interaction.
\end{enumerate}

To conclude, we would like to stress that by exploiting the universal factorization property exhibited by the short-range behavior  of the
nuclear wave function
for finite nuclei, we have  generated a microscopic and parameter-free  spectral function
 based upon  {\it ab initio} relative and center-of-mass
two-nucleon momentum distributions  for a given nucleus. The model rigorously satisfies the conditions for its validity, in that  it takes
into account only those two- and three-nucleon configurations compatible with the requirement of wave function factorization. We have
tested the convolution formula by a comparison with available {\it ab-initio} spectral functions for the three-nucleon system resulting
from the non-relativistic Hamiltonian containing realistic local two-nucleon interactions
 (Argonne AV18), finding   an excellent  agreement
  in a wide range of removal energy and
 momentum,  provided  the effects of 3N SRC are also taken into account.
It is
 highly satisfactory that such an agreement has been obtained without the use of
 any adjustable parameter.
 The generalization of our approach to complex nuclei, for which {\it ab-initio}
spectral functions cannot yet be obtained, is straightforward and will be presented elsewhere. We consider such a generalization
 particularly
useful whenever
precise calculations of nuclear effects in various processes, e.g. electron and neutrino scattering, is required. Needless to say
that these type of processes require the inclusion of all types of final-state interaction
which are at work when the active (struck) nucleon leaves the nucleus
interacting with the spectator particles.

\newpage
\newpage
\newpage
\newpage
\newpage
\newpage
\end{document}